\documentclass[reqno,12pt]{article}   %  ,a4paper
\usepackage{amssymb}
\usepackage{amsmath}
\usepackage{graphicx}
\usepackage[square,numbers]{natbib}
\usepackage{url}
\usepackage{subfigure}
\usepackage{float}

\newtheorem{theorem}{Theorem}

\newtheorem{lemma}[theorem]{Lemma}

\def\sign{\operatorname{sign}}
\def\diag{\operatorname{diag}}
\def\disk{\operatorname{disk}}

\begin{document}%%%%%%%%%%%%%%%%%%%%%%%%
\def\singlespacing{\baselineskip=13pt}	\def\doublespacing{\baselineskip=18pt}
\singlespacing%\doublespacing

\pagestyle{myheadings}\markright{{\sc  Matrix sign function for surface waves }  }

\title{The matrix sign function for solving surface wave problems in
homogeneous and laterally periodic elastic half-spaces }
\author{A.N. Norris$^{a}$, A.L. Shuvalov$^{b}$, A.A. Kutsenko$^{b}$ \\
%EndAName
\\
$^{a}$\ Mechanical and Aerospace Engineering, Rutgers University,\\
Piscataway, NJ 08854-8058, USA \\
$^{b}$ Institut de M\'{e}canique et d'Ing\'{e}nierie de Bordeaux, \\
Universit\'{e} de Bordeaux, UMR CNRS 5469, Talence 33405, France }
\maketitle

\begin{center}

\medskip \textit{Dedicated to Vladimir I. Alshits on the occasion of his
70th anniversary. }
\end{center}


\section*{Abstract}

The matrix sign function is shown to provide a simple and direct method to
derive some fundamental results in the theory of surface waves in
anisotropic materials. It is used to establish a shortcut to the basic
formulas of the Barnett-Lothe integral formalism and to obtain an explicit
solution of the algebraic matrix Riccati equation for the surface impedance.
The matrix sign function allows the Barnett-Lothe formalism to be readily
generalized for the problem of finding the surface wave speed in a
periodically inhomogeneous half-space with material properties that are
independent of depth. No partial wave solutions need to be found; the
surface wave dispersion equation is formulated instead in terms of blocks of
the matrix sign function of $i$ times the Stroh matrix.

\section{Introduction}

\label{sec_1}

The theory of surface waves in homogeneous anisotropic elastic half-spaces
has enjoyed remarkable progress in the 1960's and 1970's. The general
framework developed by Stroh \cite{Stroh62} for solving static and dynamic
elasticity problems has proved to be very fruitful for study of surface
waves. In a series of classic papers, e.g.\ \cite{Barnett73,Lothe76}, Barnett
and Lothe employed the framework of Stroh to develop an elegant integral
matrix formalism that underpins existence and uniqueness considerations for
surface waves and that allows one to determine the surface wave speed
without having to solve for any partial wave solutions. The Barnett-Lothe
integral formalism was quickly realized to serve as a corner stone for the
surface wave theory. On this basis, \citet{Chadwick77} provided a thorough
exposition of the complete theory for surface waves in anisotropic elastic
solids, summarizing the major developments up to that time, 1977. Later on,
a significant contribution came from Alshits. With his co-workers, he has
done much work on extending the formalism of Stroh, Barnett and Lothe to
various problems of crystal acoustics, see the bibliography in this Special
Issue. A full historical record and a broad overview of the surface wave
related phenomena may be found in \cite{Ting96,Barnett00}.

The purpose of this paper is both to present a fresh perspective on the
integral formalism for surface waves and also to provide new results,
including a generalization to laterally periodic half-spaces. The central
theme is the use of the matrix sign function which allows a quick derivation
and a clear interpretation of the integral formalism of \cite{Barnett73}.
The application of the matrix sign function in the context of the Stroh
formulation of elastodynamics was apparently first noted by \cite{Norris10}
in the course of calculation of impedance functions for a solid cylinder. We
reconsider the classical surface wave problem in terms of the matrix sign
function, showing in the process that it provides a natural solution
procedure. For instance, it is known that the surface impedance matrix $%
\mathbf{Z}$ satisfies an algebraic Riccati equation \cite{Biryukov95}, but it has
not been used to directly solve for $\mathbf{Z}$. Here we give the first
explicit solution of this Riccati equation for the impedance. Another
important attribute of the matrix sign function is that it allows the
Barnett-Lothe formalism to be readily generalized to finding the surface
wave speed in a periodically inhomogeneous half-space whose material
properties are independent of depth (i.e. a 2D laterally periodic
half-space). For this case, the present results %provide for
provide a procedure that circumvents the need for partial wave solutions.
Instead it establishes the dispersion equation in terms of the matrix sign
function which can be evaluated by one of the optional methods, in
particular in the integral form similar to the Barnett-Lothe representation
of the homogeneous case.

The outline of the paper is as follows. The surface wave problem is defined
in \S \ref{sec_2} in terms of the Stroh matrix $\mathbf{N}$. The matrix sign
function $\sign i\mathbf{N}$ is introduced and discussed in \S \ref%
{sec_3} where it is shown to supply a novel and possibly advantageous route
to the Barnett-Lothe integral formalism. 
% which play a central role in surface wave theory, among other fields.
An explicit solution of the algebraic Riccati equation for the surface
impedance matrix $\mathbf{Z}$ is derived in \S \ref{sec_4}. Application of
the matrix sign function to formulating and solving the surface wave
dispersion equation in a laterally periodic half-space is considered in \S %
\ref{sec_5}, with the numerical examples given for a bimaterial
configuration. The Appendix highlights explicit links between the sign
function and some related matrix functions.

\section{Background and problem definition}

\label{sec_2}

The equations of equilibrium for time harmonic motion (with the common
factor $\mathrm{e}^{-i\omega t}$ everywhere omitted but understood) are 
\begin{equation}
\partial _{j}\sigma _{ij}=-\rho \omega ^{2}u_{i},\quad \sigma
_{ij}=c_{ijkl}\partial _{l}u_{k}\ \ (i,j,k,l=1,2,3) ,  \label{3=1}
\end{equation}%
where $\rho $ is mass density, $c_{ijkl}$ are the elements of the elastic
tensor $\mathbf{C}$ referred to an orthonormal coordinate system, $u_{i}$
and $\sigma _{ij}$ are elements of the displacement $\mathbf{u}({\mathbf{x}}%
,t)$ and stress $\boldsymbol{\sigma }({\mathbf{x}},t)$. We first consider a
uniform elastic half-space $\mathbf{n}\cdot \mathbf{x}\geq 0$, ($|\mathbf{n}%
|=1$) with constant density and elastic moduli. Solutions are sought in the
form of surface waves propagating in the direction $\mathbf{m}$ parallel to
the free surface ($\mathbf{m}\cdot \mathbf{n}=0$, $|\mathbf{m}|=1$): 
%Denoting  the displacement ${\bf u}({\bf x},t) $ and the traction vector
%$\mathbf{n} \boldsymbol{\sigma}$,
% where  is the stress,  w
\begin{equation}
\begin{pmatrix}
\mathbf{u} \\ 
\mathbf{n}\boldsymbol{\sigma }%
\end{pmatrix}%
=%
\begin{pmatrix}
\mathbf{a}(\mathbf{n}\!\cdot \!\mathbf{x}) \\ 
-ik\mathbf{l}(\mathbf{n}\cdot \mathbf{x})%
\end{pmatrix}%
\,\mathrm{e}^{ik\mathbf{m}\cdot \mathbf{x}}.  \label{=1}
\end{equation}
The equations of equilibrium \eqref{3=1} take the form of a differential
equation for the 6-vector $\boldsymbol{\eta }$ \cite{Lothe76}, 
\begin{align}
\frac{\mathrm{d}\boldsymbol{\eta }}{\mathrm{d}y}& =ik\mathbf{N}\boldsymbol{%
\eta },\quad \text{with}\ \ \boldsymbol{\eta }(y)=%
\begin{pmatrix}
\mathbf{a}(y) \\ 
\mathbf{l}(y)%
\end{pmatrix}%
,  \label{=2} \\
\mathbf{N}& =-%
\begin{pmatrix}
(nn)^{-1}(nm) & (nn)^{-1} \\ 
(mn)(nn)^{-1}(nm)-(mm)+\rho v^{2}\mathbf{I} & (mn)(nn)^{-1}%
\end{pmatrix}%
,  \label{02-}
\end{align}%
where $v=\omega /k$, the 3$\times $3 matrix $\left( pq\right) $ has
components $(pq)_{jk}=p_{i}c_{ijkl}q_{l}$ for arbitrary vectors $\mathbf{p}$
and $\mathbf{q}$, and $\mathbf{I}$ is the identity matrix. The real-valued
Stroh matrix $\mathbf{N}$ satisfies $\mathbf{N}^{T}=\mathbf{K}\mathbf{N}%
\mathbf{K}$ where $^{T}$ indicates transpose and $\mathbf{K}$ in block
matrix form comprises zero blocks on the diagonal and identity blocks off
the diagonal. Denote the eigenvalues and eigenvectors of $i\mathbf{N}$ by $%
\lambda _{\alpha }$ and $\boldsymbol{\xi }_{\alpha }=\left( \mathbf{a}%
_{\alpha },\mathbf{l}_{\alpha }\right) ^{T}$ ($\alpha =1,...,6$), and
introduce the matrix $\mathbf{\Gamma }=\left\Vert \boldsymbol{\xi }_{1},...,%
\boldsymbol{\xi }_{6}\right\Vert $ with columns $\boldsymbol{\xi }_{\alpha
}. $ Assume in the following the normal situation where all $\lambda
_{\alpha }$ are distinct. 
%{\large (unless otherwise specified ??? WE'LL SEE... )} 
Then the above symmetry of $\mathbf{N}$ yields the
orthogonality/completeness relations in the form \cite[eqs.\ 2.8-2.12]%
{Barnett73} 
\begin{equation}
\mathbf{\Gamma }^{T}\mathbf{K\mathbf{\Gamma }}=\mathbf{I} \quad
\Leftrightarrow \quad \mathbf{\Gamma }\mathbf{K} \mathbf{\Gamma }^{T}=%
\mathbf{I},  \label{883-1}
\end{equation}%
where the normalization $\boldsymbol{\xi }_{\alpha }^{T}{\mathbf{K}}%
\boldsymbol{\xi }_{\alpha }=1$ has been adopted. Hereafter we use the same
notation $\mathbf{I}$ for the identity matrix regardless of its algebraic
dimension. %{\large (this is useful for \S 5!)}

Throughout this paper we restrict our interest to the subsonic surface waves
and thus assume that $v$ is less than the so-called limiting wave speed $%
\hat{v}$, see \cite{Lothe76,Chadwick77,Barnett00}. This implies that $%
i\lambda _{\alpha }$ are all nonzero and in pairs of complex conjugates
(denoted below by $^{\ast }$), so the set of eigenvalues and eigenvectors of 
$i\mathbf{N}$ can be split into a pair of triplets as 
\begin{equation}
\lambda _{\alpha }=-\lambda _{\alpha +3}^{\ast }\quad (\mathrm{Re}\lambda
_{\alpha }<0)\quad \Rightarrow \quad \boldsymbol{\xi }_{\alpha }^{\ast }=%
\boldsymbol{\xi }_{\alpha +3},\quad \alpha =1,2,3.  \label{883}
\end{equation}%
These two triplets are commonly referred to as physical $(\alpha =1,2,3)$
and nonphysical $(\alpha =4,5,6)$ since they define partial modes that,
respectively, decay or grow with increasing $\mathbf{n}\cdot \mathbf{x}$.
The eigenvector matrix $\mathbf{\Gamma }$ partitioned according to %
\eqref{883} is written in the block form as 
\begin{equation}
\mathbf{\Gamma }=%
\begin{pmatrix}
\mathbf{A}_{1} & \mathbf{A}_{2} \\ 
\mathbf{L}_{1} & \mathbf{L}_{2}%
\end{pmatrix}%
\ \ \text{with}\ \ \begin{aligned} \mathbf{A}_{1}&=\left\Vert
\mathbf{a}_{1},\mathbf{a}_{2},\mathbf{a}_{3}\right\Vert, \ \
\mathbf{A}_{2}=\left\Vert
\mathbf{a}_{4},\mathbf{a}_{5},\mathbf{a}_{6}\right\Vert = \mathbf{A}_{1}^*,
\\ \mathbf{L}_{1}&=\left\Vert
\mathbf{l}_{1},\mathbf{l}_{2},\mathbf{l}_{3}\right\Vert , \ \  \ \ \,
\mathbf{L}_{2}=\left\Vert
\mathbf{l}_{4},\mathbf{l}_{5},\mathbf{l}_{6}\right\Vert= \mathbf{L}_{1}^* ,
\end{aligned}  \label{-4}
\end{equation}%
where the blocks $\mathbf{A}_{1}$, $\mathbf{L}_{1}$ and $\mathbf{A}_{2}$, $%
\mathbf{L}_{2}$ describe the physical and nonphysical (decaying and growing)
wave solutions, respectively. From \eqref{883} $\mathbf{\Gamma }^{\ast }=%
\mathbf{\Gamma K}$ and so the orthogonality/completeness relations %
\eqref{883-1} may be cast in the subsonic domain to the form 
\begin{equation}
\mathbf{\Gamma }^{+}\mathbf{K\mathbf{\Gamma }}=\mathbf{K}\quad
\Leftrightarrow \quad \mathbf{\Gamma }\mathbf{K\mathbf{\Gamma }}^{+}=\mathbf{%
K},  \label{884}
\end{equation}%
where $^{+}$ denotes the Hermitian transpose. Note that the relations %
\eqref{884} admit interpretation in terms of the energy flux into the depth.

The surface wave solution comprises the decaying solutions only and
therefore must have the form 
\begin{equation}
\mathbf{a}(y)=\mathbf{A}_{1}\,\mathrm{e}^{y{\boldsymbol{\lambda }}}\,\mathbf{%
v},\quad \mathbf{l}(y)=\mathbf{L}_{1}\,\mathrm{e}^{y{\boldsymbol{\lambda }}%
}\,\mathbf{v}\ \ \text{with }\ {\boldsymbol{\lambda }}=\mathrm{diag}(\lambda
_{1},\lambda _{2},\lambda _{3}),  \label{-91}
\end{equation}
where $\mathbf{v}$ is some fixed vector. The surface wave problem for the
homogeneous medium is posed as finding for a given wavenumber $k\in \mathbb{R%
}^{+}$ the surface wave speed $v=v_{s}\in \mathbb{R}$ for which the surface
traction $\mathbf{l}(0)$ vanishes, and hence $\mathbf{v}$ is a null vector
of $\mathbf{L}_{1}$ (although this is not a fruitful avenue to follow, which
is the whole point of the Barnett-Lothe solution procedure based on the
integral representation).

A variety of related notations have been used for the surface wave problem.
We generally follow \citet{Barnett85} where the notation is based upon that
of \cite{Lothe76}. \citet{Barnett85} also provide comparisons of their
notation with that used by \citet{Chadwick77}, which is closer
to that of \citet{Ting96}. The slight notational
differences are related to the choice of the $6-$vector in \eqref{=2}, and
amounts to different signs for the diagonal or off-diagonal elements of the $%
6\times 6$ matrix analogous to $\mathbf{N}$.

\section{The matrix sign function $\sign i\mathbf{N}$}

\label{sec_3}

\subsection{Definition}

\label{3.1}

The sign function of a matrix $\mathbf{M}$ is conveniently defined by
analogy with the scalar definition as \cite{Higham:1994:MSD} 
\begin{equation}
\sign\mathbf{M}=\mathbf{M}\,\big(\mathbf{M}^{2}\big)^{-1/2},
\label{273}
\end{equation}%
where the principle branch of the square root function with branch cut on
the negative real axis is understood; $z=(z^{2})^{1/2}\,\sign z$ with $%
\sign z=+1(-1)$ if $\mathrm{Re}z>0(<0)$. As a result, the sign function
of a matrix $\mathbf{M}$ with eigenvalues and eigenvectors denoted by $%
\lambda _{\alpha }$ and $\boldsymbol{\xi }_{\alpha }$ satisfies 
\begin{equation}
\mathbf{M}\boldsymbol{\xi }_{\alpha }=\lambda _{\alpha }\boldsymbol{\xi }%
_{\alpha }\quad \Rightarrow \quad \big(\sign\mathbf{M}\big)\boldsymbol{%
\xi }_{\alpha }=\pm \boldsymbol{\xi }_{\alpha }\ \mathrm{for}\ \lambda
_{\alpha }\in \mathbb{C}^{\pm }.  \label{274}
\end{equation}%
Note that $\sign\mathbf{M}$ is unchanged under $\mathbf{M}\rightarrow a%
\mathbf{M}+ib\mathbf{I}$, $a\in \mathbb{R}^{+},\,b\in \mathbb{R}$, and it is
undefined for eigenvalues lying on the imaginary axis ($\mathrm{Re}\lambda
_{\alpha }=0$). We also note for later use the property $\sign\big(%
\mathbf{U}\mathbf{M}\mathbf{U}^{-1}\big)=\mathbf{U}\big(\sign\mathbf{M}%
\big)\mathbf{U}^{-1}$. Evaluation of the matrix sign function is possible
with a variety of numerical methods \cite{Higham:2008:FM}, the simplest
being Newton iteration of $\mathbf{M}_{k+1}=\frac{1}{2}(\mathbf{M}_{k}+%
\mathbf{M}_{k}^{-1})$, $\mathbf{M}_{0}=\mathbf{M}$, with $\mathbf{M}%
_{k}\rightarrow \sign\mathbf{M}$ in the limit as $k\rightarrow \infty $%
, although this can display convergence problems. Schur decomposition, which
does not require matrix inversion, is very stable, and is readily available,
e.g.\ \cite{Higham}. The function $\sign\mathbf{M}$ also has integral
representations, which we will use in order to shed fresh light on the
integral formalism in the surface wave theory. See \cite%
{Kenney:1995:MSF,Higham:2008:FM} for reviews of  the matrix sign function.

The following expression for the matrix sign function is based on \eqref{273}
combined with an integral representation for the matrix square root function 
\cite{Roberts:1980:LMR,Higham:2008:FM} 
\begin{equation}
\sign\mathbf{M}=\frac{2}{\pi }\mathbf{M}\int_{0}^{\infty }\mathrm{d}t\,%
\big(t^{2}\mathbf{I}+\mathbf{M}^{2}\big)^{-1}.  \label{33.21}
\end{equation}%
Equation \eqref{33.21} may be converted into the following form using a
change of integration variable 
\begin{equation}
\sign\mathbf{M}=\left\langle \mathbf{M}_{\theta }\right\rangle \ \ 
\text{with }\ \ \mathbf{M}_{\theta }\equiv (\cos \theta \mathbf{I}-i\sin
\theta \mathbf{M})^{-1}(\cos \theta \mathbf{M}-i\sin \theta \mathbf{I}),
\label{-55-}
\end{equation}%
where 
\begin{equation*}
\langle \cdot \rangle =\frac{1}{\pi }\int_{0}^{\pi }\cdot \,\mathrm{d}\theta
\end{equation*}%
denotes the average. Differentiation of $\mathbf{M}_{\theta }$ yields 
\begin{equation}
\frac{\mathrm{d}\mathbf{M}_{\theta }}{\mathrm{d}\theta }=i\big(\mathbf{M}%
_{\theta }^{2}-\mathbf{I}\big),  \label{000}
\end{equation}%
which implies $\langle \mathrm{e}^{i\phi \mathbf{M}_{\theta }}\rangle
=\langle \mathrm{e}^{-i\phi \mathbf{M}_{\theta }}\rangle ^{-1}=\cos \phi 
\mathbf{I}+i\sin \phi \langle \mathbf{M}_{\theta }\rangle $ for any $\phi $.
This provides alternative identities for the matrix sign function, such as 
\begin{equation}
\sign\mathbf{M}=\big\langle\frac{\sin (\phi \mathbf{M}_{\theta })}{%
\sin \phi }\big\rangle=\big\langle\frac{\mathbf{M}_{\theta }\cos (\phi 
\mathbf{M}_{\theta })}{\cos \phi }\big\rangle,  \label{-9}
\end{equation}%
where the value of $\phi $ is arbitrary as long as the denominator, $\sin
\phi $ or $\cos \phi $, does not vanish. The special case of eq.\ %
\eqref{-55-}$_{1}$ corresponds to the limit as $\phi \rightarrow 0$ in %
\eqref{-9}.

\subsection{ $\sign i\mathbf{N}$ and the
Barnett-Lothe matrices}

\label{3.2}

Let us apply the above definition and properties of the matrix sign function
to the case $\sign i\mathbf{N}$ where $\mathbf{N}$ is the Stroh matrix
given in \eqref{02-} and considered in the subsonic domain $v<\hat{v}$. The
eigenspectrum of $\mathbf{N}$ is assumed partitioned according to (\ref{883}%
). Hence equation (\ref{274})$_{2}$ taken with $\mathbf{M}=i\mathbf{N}$ reads%
\begin{equation}
\left( \sign i\mathbf{N}\right) \boldsymbol{\xi }_{\alpha }=-%
\boldsymbol{\xi }_{\alpha },\ \left( \sign i\mathbf{N}\right) 
\boldsymbol{\xi }_{\alpha +3}=\boldsymbol{\xi }_{\alpha +3},\ \alpha =1,2,3,
\label{eigen}
\end{equation}%
Denote the blocks of $\sign i\mathbf{N\ }$\textbf{(}$=\mathbf{K}\left( 
\sign i\mathbf{N}\right) ^{T}\mathbf{K}$) as 
\begin{equation}
\sign i\mathbf{N}=i%
\begin{pmatrix}
\mathbf{S} & \mathbf{Q} \\ 
\mathbf{B} & \mathbf{S}^{T}%
\end{pmatrix}%
\ \mathrm{with}\ \mathbf{Q=Q}^{T},\ \mathbf{B=B}^{T},\ \text{tr}\,\mathbf{S}%
=0,  \label{def}
\end{equation}%
where $\mathbf{S,\ Q,\ B}$ are all real since so is $\mathbf{N}$. Note that $%
\sign i\mathbf{N}$ and hence $\mathbf{S}$ are traceless. 
%The matrix $\mathbf{Q}$ is negative definite due to {\large HOW TO\ SHOW\ with no appealto }$\int ${\large \ representation?} 
Appearance of the Barnett-Lothe notations on the right-hand side will become
clear in the course of the upcoming derivation. The involutory property of
the matrix sign function, $\big(\sign i\mathbf{N}\big)^{2}=\mathbf{I}$,
implies the identities 
\begin{equation}
\mathbf{BS}+\mathbf{S}^{T}\mathbf{B}=\mathbf{0},\ \ \mathbf{SQ}+\mathbf{QS}%
^{T}=\mathbf{0},\ \ \mathbf{S}^{2}+\mathbf{QB}=-\mathbf{I}.  \label{7.16}
\end{equation}%
Taking into account the spectral representation $i\mathbf{N}=\mathbf{\Gamma }%
\mathrm{diag}(\boldsymbol{\lambda },-\boldsymbol{\lambda }^{\ast })\mathbf{%
\Gamma }^{-1}$ and the relation (\ref{883-1})$_{2}$ 
%{\large (I suggest to stay real so far...)} 
yields the spectral decomposition of the matrix $\sign i\mathbf{N}$ in
the form%
\begin{equation}
\begin{aligned} \sign i\mathbf{N} &= \mathbf{\Gamma
}\mathrm{diag}(-\mathbf{I},\mathbf{I})\mathbf{\Gamma
}^{T}\mathbf{K}=\begin{pmatrix}
\mathbf{A}_{2}\mathbf{L}_{2}^{T}-\mathbf{A}_{1}\mathbf{L}_{1}^{T} &
\mathbf{A}_{2}\mathbf{A}_{2}^{T}-\mathbf{A}_{1}\mathbf{A}_{1}^{T} \\
\mathbf{L}_{2}\mathbf{L}_{2}^{T}-\mathbf{L}_{1}\mathbf{L}_{1}^{T} &
\mathbf{L}_{2}\mathbf{A}_{2}^{T}-\mathbf{L}_{1}\mathbf{A}_{1}^{T}%
\end{pmatrix} \\ & =\mathbf{I}-2\begin{pmatrix}
\mathbf{A}_{1}\mathbf{L}_{1}^{T} & \mathbf{A}_{1}\mathbf{A}_{1}^{T} \\
\mathbf{L}_{1}\mathbf{L}_{1}^{T} &
\mathbf{L}_{1}\mathbf{A}_{1}^{T}\end{pmatrix}=\mathbf{I}-2 \begin{pmatrix}
\mathbf{A}_{1} \\ \mathbf{L}_{1}\end{pmatrix}\begin{pmatrix} \mathbf{A}_{1}
\\ \mathbf{L}_{1}\end{pmatrix}^{T} \mathbf{K\ }\left(
=\mathbf{I}-2\mathbf{P}_{-}\right) ,\mathbf{\ \ }\end{aligned}
\label{spectral}
\end{equation}%
where we have noted the projector $\mathbf{P}_{-}$ on physical modes: $%
\mathbf{P}_{-}\mathbf{\Gamma =\Gamma }${\scriptsize {$%\footnotesize
\begin{pmatrix}
\mathbf{I} & \mathbf{0} \\ 
\mathbf{0} & \mathbf{0}%
\end{pmatrix}%
$}} (see more in Appendix). From (\ref{def}) and (\ref{spectral}),%
\begin{equation}
\mathbf{S}=i(2\mathbf{A}_{1}\mathbf{L}_{1}^{T}-\mathbf{I}),\ \mathbf{Q}=2i%
\mathbf{A}_{1}\mathbf{A}_{1}^{T},\quad \mathbf{B}=2i\mathbf{L}_{1}\mathbf{L}%
_{1}^{T}.  \label{==3}
\end{equation}

Assume that $\mathbf{A}_{1}$ and hence $\mathbf{Q}$ are invertible (we will
return to this point later). Introduce the surface impedance $\mathbf{Z}=i%
\mathbf{L}_{1}\mathbf{A}_{1}^{-1}$. It is Hermitian due to (\ref{884})$_{1}.$
Plugging (\ref{def}) into (\ref{eigen})$_{1}$ gives $\mathbf{SA}_{1}+\mathbf{%
QL}_{1}=i\mathbf{A}_{1}$ which, with regard for the invertibility of $%
\mathbf{Q}$, enables expressing $\mathbf{Z}$ through the blocks of $\sign i{\mathbf{N.}}$ Thus%
\begin{equation}
\mathbf{Z}=i\mathbf{L}_{1}\mathbf{A}_{1}^{-1}\left( =\mathbf{Z}^{+}\right) \
\Leftrightarrow \ \mathbf{Z}=-\mathbf{Q}^{-1}(\mathbf{I}+i\mathbf{S})~.
\label{7.22}
\end{equation}%
Note from (\ref{7.22}) that, for any $v,$ $\det \mathbf{S}=0$ (since $%
\mathbf{Q}^{-1}\mathbf{S}$ is antisymmetric, see also (\ref{7.16})$_{2}$)
and that $\det \left( \mathbf{I}+i\mathbf{S}\right) $ is real (since $%
\mathbf{Z=Z}^{+}$).

General significance of $\sign i\mathbf{N}$ for surface waves is
immediate when one considers the boundary condition which demands that a
non-zero surface displacement exerts zero surface traction. Since a surface
wave is composed of decaying modes only, this means there must exist a
vector $\mathbf{v}\neq 0$ such that%
\begin{equation}
\left( \sign i{\mathbf{N}}\right) 
\begin{pmatrix}
\mathbf{v} \\ 
\mathbf{0}%
\end{pmatrix}%
=%
\begin{pmatrix}
-\mathbf{v} \\ 
\mathbf{0}%
\end{pmatrix}%
\quad \Rightarrow \quad \left\{ \begin{aligned} ({\mathbf I} +i {\mathbf S}
) {\bf v} &= 0, \\ {\mathbf B}{\bf v} &=0 . \end{aligned}\right.  \label{-0}
\end{equation}%
Hence both $\det \mathbf{B}$ and $\det \left( \mathbf{I}+i\mathbf{S}\right) $
must vanish at the surface wave speed $v=v_{s}$. For the case in hand of a
uniform material, these are not two independent conditions but one
condition. Indeed, the traction-free boundary condition can be posed in the
form $\det \mathbf{L}_{1}=0,$ hence from (\ref{==3}) and (\ref{7.22})$_{1}$
determinants of $\mathbf{B,}$ of $\mathbf{I}+i\mathbf{S~}\left( =2\mathbf{A}%
_{2}\mathbf{L}_{1}^{+}\right) $ and of $\mathbf{Z}$ vanish simultaneously.
Note that the equation $\det \left( \mathbf{I}+i\mathbf{S}\right) =0$
reduces to $\mathrm{tr}\left( {\mathbf{S}}^{2}\right) =-2$ (since $\mathrm{tr%
}\mathbf{S}=0$ and $\det \mathbf{S}=0$). {Also at }$v=v_{s},$ the common
null vector $\mathbf{v}$ of $\mathbf{I}+i\mathbf{S}$ and $\mathbf{B}$ can be
cast as $\mathbf{v}=\mathbf{a}+i\mathbf{b}$ with linear independent real
vectors $\mathbf{a}$ and $\mathbf{b,}$ hence $\mathbf{B}$ is rank one. To
summarize{, the subsonic surface wave speed }$v=v_{s}${\ can be determined
from }$\det \mathbf{L}_{1}=0${\ or else from any of the equivalent real
dispersion equations (see \cite[eqs.\ (12.10-1)-(12.10-4)]{Ting96}) 
\begin{equation}
\det \mathbf{Z}=0\ \Leftrightarrow \mathrm{tr}\left( {\mathbf{S}}^{2}\right)
=-2\ \Leftrightarrow \det {\mathbf{B}}=0\ \ \left( \Rightarrow {(\mathrm{tr}{%
\mathbf{B}})^{2}=\mathrm{tr}({\mathbf{B}}^{2})}\right) ,  \label{36}
\end{equation}%
which all are expressed through the blocks of the matrix }$\sign i{%
\mathbf{N}}$.

Equations (\ref{7.16}), (\ref{==3}), (\ref{7.22}), (\ref{36}) are basic
relations of the Barnett-Lothe formalism with regard to surface wave theory. 
%{\large (leaving aside the sign-definiteness properties =\TEXTsymbol{>} in Conclusion?)}. 
Interestingly, we have arrived at these relations by single means of the
definition of sign function of  $i$ times the Stroh matrix $\mathbf{N}=%
\mathbf{K}\mathbf{N}^{T}\mathbf{K}$, without specifying the method of this
function evaluation and without explicitly attending to the fundamental
elasticity tensor which underlies development of Barnett-Lothe formalism in 
\cite{Barnett73,Lothe76,Chadwick77}. It is instructive to highlight an
explicit link between the two lines of derivation, see next.

\subsection{Linking $\sign i\mathbf{N}$ to the fundamental
elasticity tensor $\mathbf{N}_{\protect\theta }$}

Following \cite{Barnett73,Lothe76,Chadwick77}, introduce the so-called \cite%
{Chadwick77} fundamental elasticity tensor 
\begin{equation}
\begin{aligned} & \mathbf{N}_{\theta }= -\begin{pmatrix} \lbrack
\![ss]\!]^{-1}[\![sr]\!] & [\![ss]\!]^{-1} \\[0pt] \lbrack
\![rs]\!][\![ss]\!]^{-1}\![\![sr]\!]-[\![rr]\!] &
[\![rs]\!][\![ss]\!]^{-1}\end{pmatrix} \\ & \mathrm{where}\
[\![pq]\!]=\left( pq\right) -\left[ \mathbf{m\cdot p}\right] \left[
\mathbf{m\cdot q}\right] \rho v^{2}\mathbf{I\ \ } \mathrm{with}\ 
\mathbf{p,q}=\mathbf{r},\mathbf{s}: \\ & \qquad \qquad \mathbf{r}=\cos
\theta \mathbf{m}+\sin \mathbf{n},\ \mathbf{s=-}\sin \theta \mathbf{m+}\cos
\theta \mathbf{n.}\end{aligned}  \label{fund}
\end{equation}%
The matrix $\mathbf{N}_{\theta }$ is defined in the subsonic domain which
means that $v$ is small enough to guarantee existence of $[\![ss]\!]^{-1}$
for any $\theta $. Note that $\mathbf{N}_{\theta }$ at $\theta =0$ is equal
to the Stroh matrix $\mathbf{N}$ defined in (\ref{02-}).

\begin{lemma}
\label{-11} The matrix sign function of $i$ times the Stroh matrix $\mathbf{N%
}$ is $i$ times the angular average of the fundamental elasticity tensor $%
\mathbf{N}_{\theta }$ 
\begin{equation}
\sign i\mathbf{N}=i\langle \mathbf{N}_{\theta }\rangle \ \Rightarrow \ 
\begin{array}{c}
\mathbf{S}=-\left\langle [\![ss]\!]^{-1}[\![sr]\!]\right\rangle ,\ \mathbf{Q}%
=-\left\langle [\![ss]\!]^{-1}\right\rangle \\ 
\mathbf{B}=\left\langle [\![rr]\!] - [\![rs]\!][\![ss]\!]^{-1}
[\![sr]\!]\right\rangle .%
\end{array}
\label{82}
\end{equation}
\end{lemma}

Proof: Equation (\ref{-55-}) with $\mathbf{M}\equiv i\mathbf{N}$ 
%and $\mathbf{M}_{\theta }\equiv i\mathbf{G}$ 
reads 
\begin{equation}
\sign i\mathbf{N}=i\langle \mathbf{G}\rangle \quad \text{with}\ \ 
\mathbf{G}\equiv (\cos \theta \mathbf{I}+\sin \theta \mathbf{N})^{-1}(\cos
\theta \mathbf{N}-\sin \theta \mathbf{I}).  \label{=3}
\end{equation}%
Let us show that $\mathbf{N}_\theta $ defined in (\ref{fund})
and $\mathbf{G}$ given by (\ref{=3})$_{2}$ are one and the same.
Substitution into $\mathbf{G}$ the representation \cite[eq.\ (3.15)]%
{Chadwick77} 
\begin{equation}
\begin{aligned} & \mathbf{N} = (\mathbf{X} - \mathbf{Y})^{-1} \mathbf{J}
(\mathbf{X} + \mathbf{Y}) \ \ \text{where} \ \ \mathbf{J} = \begin{pmatrix}
\mathbf{0} & \mathbf{I} \\ - \mathbf{I} & \mathbf{0}\end{pmatrix} , \\ &
\mathbf{X} - \mathbf{Y} = \begin{pmatrix} (n n)& {\bf 0} \\ (m n) & -
\mathbf{I} \end{pmatrix} , \ \ \mathbf{X} + \mathbf{Y} = \begin{pmatrix}
[\![mm]\!]& {\bf 0} \\ -(nm) & - \mathbf{I} \end{pmatrix}, \end{aligned}
\label{=22}
\end{equation}%
yields, using $\mathrm{e}^{\mathbf{J}\theta }=\cos \theta \mathbf{I}+\sin
\theta \mathbf{J}$, 
\begin{align}
\mathbf{G}=& \big(\cos \theta (\mathbf{X}-\mathbf{Y})+\sin \theta \mathbf{J}(%
\mathbf{X}+\mathbf{Y})\big)^{-1}\big(\cos \theta \mathbf{J}(\mathbf{X}+%
\mathbf{Y})-\sin \theta (\mathbf{X}-\mathbf{Y})\big)  \notag  \label{3-9} \\
=& \big(\mathrm{e}^{\mathbf{J}\theta }\mathbf{X}-\mathrm{e}^{-\mathbf{J}%
\theta }\mathbf{Y}\big)^{-1}\mathbf{J}\big(\mathrm{e}^{\mathbf{J}\theta }%
\mathbf{X}+\mathrm{e}^{-\mathbf{J}\theta }\mathbf{Y}\big)  \notag \\
=& \big(\mathbf{X}-\mathrm{e}^{-2\mathbf{J}\theta }\mathbf{Y}\big)^{-1}%
\mathbf{J}\big(\mathbf{X}+\mathrm{e}^{-2\mathbf{J}\theta }\mathbf{Y}\big).
\end{align}%
Noting the identities 
\begin{equation}
\mathbf{X}-\mathrm{e}^{-2\mathbf{J}\theta }\mathbf{Y}=%
\begin{pmatrix}
\lbrack \![ss]\!] & \mathbf{0} \\[0pt] 
\![\![rs]\!] & -\mathbf{I}%
\end{pmatrix}%
,\ \ \mathbf{X}+\mathrm{e}^{-2\mathbf{J}\theta }\mathbf{Y}=%
\begin{pmatrix}
\lbrack \![rr]\!] & \mathbf{0} \\ 
-[\![sr]\!] & -\mathbf{I}%
\end{pmatrix}%
,  \label{7.12}
\end{equation}%
it follows that 
\begin{equation}
\mathbf{G}=%
\begin{pmatrix}
\lbrack \![ss]\!] & \mathbf{0} \\[0pt] 
\![\![rs]\!] & -\mathbf{I}%
\end{pmatrix}%
^{-1}%
\begin{pmatrix}
\mathbf{0} & \mathbf{I} \\ 
-\mathbf{I} & \mathbf{0}%
\end{pmatrix}%
\begin{pmatrix}
\lbrack \![rr]\!] & \mathbf{0} \\ 
-[\![sr]\!] & -\mathbf{I}%
\end{pmatrix}%
=\mathbf{N}_{\theta }.  \label{7.122}
\end{equation}%
Combining (\ref{7.122}) with (\ref{=3})$_{1}$ completes the proof of (\ref%
{82}). $\blacksquare $

Note that by the definition of $\mathbf{G}$ in \eqref{=3} the eigenvectors
of $\mathbf{N}_{\theta }(=\mathbf{G})$ are independent of $\theta $ and that
eq.\ (\ref{000}) with $\mathbf{M}_{\theta }\equiv i\mathbf{\mathbf{N}}%
_{\theta }$ reads%
\begin{equation}
\frac{\mathrm{d}\mathbf{N}_{\theta }}{\mathrm{d}\theta }=-\mathbf{I}-\mathbf{%
N}_{\theta }^{2}.  \label{diff}
\end{equation}
%An essential remark concerns existence of $\mathbf{Q}$ ... {\large and its
%invertibility ... STUCK HERE ... Play down our "independence" of }$\int $%
%{\large \ formalism which is trumpeted below ?? NOTE that invertibility of }$%
%\mathbf{Q~\Leftrightarrow A}$ {\large matters for }$\mathbf{Z}${\large \ only%
%}.

Thus we have used the defining properties of sign function of the Stroh
matrix $\mathbf{N}$ to obtain the basic relations (\ref{7.16}), (\ref{==3}),
(\ref{7.22}), (\ref{36}) of the Barnett-Lothe integral formalism, and then
we made appear the fundamental elasticity matrix $\mathbf{N}_{\theta }$ but
only in the context of one of possible (integral) representations of the
sign function. This is different from the original methodology of \cite%
{Barnett73,Lothe76,Chadwick77} which follows a somewhat inverse line of
derivation. It proceeds from $\mathbf{N}_{\theta }$ and establishes its
properties such as (\ref{diff}) which are then used to arrive at the
relations (\ref{7.16}), (\ref{==3}), (\ref{7.22}), (\ref{36}). One way or
another, the core is that the surface wave speed $v=v_{s}$ can be defined
through the matrix sign function which admits evaluation by different
integral formulas, e.g.\ eqs.\ \eqref{33.21}, \eqref{-9}, \eqref{-7} and %
\eqref{2}, or iteration schemes, or otherwise. Interestingly, some of these
'indirect' methods for calculating $\langle \mathbf{N}_{\theta }\rangle $
have been discussed in the literature, though with no mention of the matrix sign
function. For instance, \citet{Gundersen87} propose a scheme tantamount to
Newton iteration, while \citet{Condat87} describe a method based on
continued fractions that appears to be similar to a stable algorithm for
computing the matrix sign function \cite{Koc94}.

In conclusion, a remark is in order concerning invertibility of $%
\mathbf{Q}$ in the subsonic domain $v<\widehat{v}$. By definition of the
latter, the eigenvalues $\mathbf{N}$ for  $v<\widehat{v}$ cannot lie on the
real axis, so $\mathbf{G}$ in (\ref{=3})$_{2}$ is well-defined and hence,
due to $\mathbf{G=N}_{\theta },$ the matrix $[[ss]]$ in (\ref{fund}) is
non-singular. Since $[[ss]]$ is obviously positive definite at $v=0,$ it
follows that $[[ss]]$ is positive within the whole subsonic domain $v<%
\widehat{v}$ (this of itself is an alternative definition of the subsonic
domain). Using $\mathbf{Q=-}\left\langle [[ss]]^{-1}\right\rangle $ as
given in (\ref{82}) shows that $\mathbf{Q}$ is indeed invertible in the
subsonic domain which  implies that $\mathbf{A}_{1,2}$ are invertible
and $\mathbf{Z}$ exists at any $v<\widehat{v}$. 
Note that we did not employ the fundamental elasticity tensor $\mathbf{N}_{\theta }$ in this proof.

\section{The surface impedance as defined by Riccati equation}

\label{sec_4}   

The surface impedance matrix $\mathbf{Z,}$ which is introduced in the
subsonic domain by any of the following relations 
\begin{equation}
\mathbf{l}_{\alpha }=-i\mathbf{Z}\mathbf{a}_{\alpha }\ \ \left( \alpha
=1,2,3\right) \quad \Leftrightarrow \quad \mathbf{l}\left( y\right) =-i%
\mathbf{Z}\mathbf{a}\left( y\right) \quad \Leftrightarrow \quad \mathbf{Z}=i%
\mathbf{L}_{1}\mathbf{A}_{1}^{-1},  \label{2222}
\end{equation}%
plays a central role in the theory of surface waves. It is a crucial
ingredient in   surface-wave uniqueness and  existence considerations 
 \cite{Lothe76,Barnett85} and is of fundamental significance for energy and flux  \cite{Ingebrigtsen69}. Direct use of the impedance has
found wide application, including guided waves and scattering in
multilayered structures, e.g.\ \cite{Honein91,Wang02,Shuvalov02,Hosten03}, edge waves in
anisotropic plates and cylindrical shells \cite{Fu03,Fu12}, and 
%surface waves in prestressed elastic solids \cite{Fu09},  
waves in cylindrically anisotropic solids \cite{Norris10}. The surface
impedance $\mathbf{Z}$ has been shown in \S \ref{3.2} to satisfy the
Barnett-Lothe formula (\ref{7.22}) which can be evaluated by integral
representation or other optional expressions for $\sign i\mathbf{N}$.
Alternatively, the impedance can be defined via differential and algebraic
Riccati equations. Thus, Biryukov \cite{Biryukov85,Biryukov95} developed a
general impedance approach for surface waves in inhomogeneous half-spaces
based on the differential Riccati equation, see also \cite{Caviglia02}. For
the homogeneous half-space considered here, substitution of $\mathbf{l}=-i%
\mathbf{Z}\mathbf{a}$ in eq.\ \eqref{=2} combined with the fact that the
impedance $\mathbf{Z}$ associated with the surface wave solution \eqref{-91}
is a constant matrix independent of $y$ leads to the algebraic Riccati
equation \cite[eq.\ 3.3.7]{Biryukov95} 
\begin{equation}
\big(\mathbf{Z}-i(mn)\big)(nn)^{-1}\big(\mathbf{Z}+i(nm)\big)-(mm)+\rho v^{2}%
\mathbf{I}=0.  \label{6-7}
\end{equation}%
This equation has a unique Hermitian solution for $\mathbf{Z}$ which is
positive definite for $v<v_{s}$ \cite{Fu02}. It is clear that the matrix $%
\mathbf{Z}$ of the Riccati equation \eqref{6-7} must be identical to $%
\mathbf{Z}$ defined within the Barnett-Lothe formalism, see \cite%
{Fu02}; however, no explicit solution of the algebraic Riccati
equation has been provided. Here we derive such a solution to \eqref{6-7}
using a method based on the matrix sign function (in fact it is for this
class of matrix problems that the sign function was first proposed \cite%
{Roberts:1980:LMR}).  

The main idea of this method is to rewrite the Riccati equation in a form
where the matrix sign function reduces it to a linear equation for the
unknown matrix:

\begin{lemma}
\label{2=2} For some matrix $\mathbf{R}$ the algebraic Riccati equation %
\eqref{6-7} is equivalent to the following identity, 
\begin{equation}
\sign i\mathbf{N}=%
\begin{pmatrix}
-\mathbf{I} & \mathbf{I} \\ 
i\mathbf{Z} & \mathbf{0}
\end{pmatrix}%
\begin{pmatrix}
-\mathbf{I} & \mathbf{R} \\ 
\ \mathbf{0} & \mathbf{I}%
\end{pmatrix}%
\begin{pmatrix}
-\mathbf{I} & \mathbf{I} \\ 
i\mathbf{Z} & \mathbf{0}%
\end{pmatrix}%
^{-1}.  \label{-84}
\end{equation}
\end{lemma}

Substitution from (\ref{def})$_{1}$ into \eqref{-84} yields 
\begin{equation}
i%
\begin{pmatrix}
\mathbf{S} & \mathbf{Q} \\ 
\mathbf{B} & \mathbf{S}^{T}%
\end{pmatrix}%
\begin{pmatrix}
-\mathbf{I} & \mathbf{I} \\ 
i\mathbf{Z} & \mathbf{0}%
\end{pmatrix}%
=%
\begin{pmatrix}
-\mathbf{I} & \mathbf{I} \\ 
i\mathbf{Z} & \mathbf{0}%
\end{pmatrix}%
\begin{pmatrix}
-\mathbf{I} & \mathbf{R} \\ 
\ \mathbf{0} & \mathbf{I}%
\end{pmatrix}%
,  \label{-85}
\end{equation}%
hence, \textit{inter alia} $\mathbf{R}=\mathbf{I}-i\mathbf{S}$, and more
importantly 
\begin{equation}
\mathbf{Z=}-\mathbf{Q}^{-1}(\mathbf{I}+i\mathbf{S}),\ \ \mathbf{B}=\left( 
\mathbf{I}+i\mathbf{S}^{T}\right) \mathbf{Z=Z}\left( \mathbf{I}-i\mathbf{S}%
\right)  \label{-86}
\end{equation}%
where (\ref{-86})$_{1}$ is the same as (\ref{7.22}) and (\ref{-86})$_{2,3}$
follow from combining identities (\ref{eigen})$_{1}$ and (\ref{7.16}) (note
that $\mathbf{I}+i\mathbf{S}^{T}$ and $\mathbf{I}-i\mathbf{S}$ are not
invertible at the surface wave speed $v=v_{s}$). We have therefore
constructed an explicit solution of the algebraic Riccati equation %
\eqref{6-7} for the surface impedance tensor $\mathbf{Z}$ and shown that
this solution coincides with the Barnett-Lothe expression (\ref{7.22}). It
remains to justify the claimed identity of Lemma \ref{2=2}.

Proof of Lemma \ref{2=2}: Consider the $6\times 6$ matrix equality 
\begin{equation}
i\mathbf{N}=i%
\begin{pmatrix}
\mathbf{N}_{1} & \mathbf{N}_{2} \\ 
\mathbf{N}_{3} & \mathbf{N}_{1}^{T}%
\end{pmatrix}%
=%
\begin{pmatrix}
-\mathbf{I} & \mathbf{I} \\ 
i\mathbf{Z} & \mathbf{0}%
\end{pmatrix}%
\begin{pmatrix}
\mathbf{M}_{1} & {\mathbf{M}_{0}} \\ 
\mathbf{0} & \mathbf{M}_{2}%
\end{pmatrix}%
\begin{pmatrix}
-\mathbf{I} & \mathbf{I} \\ 
i\mathbf{Z} & \mathbf{0}%
\end{pmatrix}%
^{-1},  \label{-83}
\end{equation}%
where 
\begin{equation}
{\mathbf{M}_{0}=\mathbf{Z}^{-1}\mathbf{N}_{3},}\quad \mathbf{M}_{1}=-\mathbf{%
Z}^{-1}(\mathbf{N}_{3}+i\mathbf{N}_{1}^{T}\mathbf{Z}),\quad \mathbf{M}_{2}=%
\mathbf{Z}^{-1}(\mathbf{N}_{3}-i\mathbf{Z}\mathbf{N}_{1}).  \label{2=1}
\end{equation}%
Simple algebra shows that the matrix Riccati equation \eqref{6-7} is
identical to the upper right block in \eqref{-83} while the other blocks are
trivial identities. Referring to the matrices $\mathbf{M}_{1}$ and $\mathbf{M%
}_{2}$, we note that the eigensolutions $\lambda _{\alpha }$ and $%
\boldsymbol{\xi }_{\alpha }=\left( \mathbf{a}_{\alpha },\mathbf{l}_{\alpha
}\right) ^{\mathrm{T}}$ $(\alpha =1,2,3)$ satisfy, from eq.\ \eqref{=2}, 
\begin{equation}
-\mathbf{N}_{3}{\mathbf{a}}_{\alpha }+\mathbf{N}_{1}^{T}{\mathbf{l}}_{\alpha
}=i\lambda _{\alpha }{\mathbf{l}}_{\alpha }\ \ \Rightarrow \ \ -\mathbf{N}%
_{3}{\mathbf{A}}_{1}+\mathbf{N}_{1}^{T}{\mathbf{L}}_{1}=i{\mathbf{L}}_{1}%
\boldsymbol{\lambda }.  \label{2=4}
\end{equation}%
Replacing ${\mathbf{L}}_{1}=-i{\mathbf{Z}}{\mathbf{A}}_{1}$ in \eqref{2=4}
yields $\mathbf{M}_{1}$ while $\mathbf{M}_{2}$ follows similarly from the
substitution ${\mathbf{A}}_{1} = i {\mathbf{Z}}^{-1} {\mathbf{L}}_{1}$, 
\begin{equation}
\mathbf{M}_{1}={\mathbf{A}}_{1}\boldsymbol{\lambda }{\mathbf{A}}%
_{1}^{-1},\quad \mathbf{M}_{2}^{+}=-{\mathbf{L}}_{1}\boldsymbol{\lambda }{%
\mathbf{L}}_{1}^{-1}.  \label{1234}
\end{equation}%
It is clear from these representations that $\mathbf{M}_{1}$ and $\mathbf{M}%
_{2}$ have eigenvalues in $\mathbb{C}^{-}$ and $\mathbb{C}^{+}$,
respectively. Define $\mathbf{R}$ as the solution of the following {Sylvester%
} equation 
\begin{equation}
\mathbf{R}\mathbf{M}_{2}-\mathbf{M}_{1}\mathbf{R}=2{\mathbf{M}_{0}}\ \
\Rightarrow \ \ 
\begin{pmatrix}
\mathbf{I} & \frac12 \mathbf{R} \\ 
\mathbf{0} & \mathbf{I}%
\end{pmatrix}%
^{-1}%
\begin{pmatrix}
\mathbf{M}_{1} & {\mathbf{M}_{0}} \\ 
\mathbf{0} & \mathbf{M}_{2}%
\end{pmatrix}%
\begin{pmatrix}
\mathbf{I} & \frac12 \mathbf{R} \\ 
\mathbf{0} & \mathbf{I}%
\end{pmatrix}%
=%
\begin{pmatrix}
\mathbf{M}_{1} & \mathbf{0} \\ 
\mathbf{0} & \mathbf{M}_{2}%
\end{pmatrix}%
,  \label{=55}
\end{equation}%
then \eqref{-84} follows from eq.\ \eqref{-83} and $\sign \diag ( \mathbf{M}_{1}, \mathbf{M}_{2}) = \diag ( -\mathbf{I}, \mathbf{I})$. 
$\blacksquare $

\section{Surface waves in a laterally periodic half-space}

\label{sec_5}

\subsection{General setup}

\label{5.1}

The equations of equilibrium \eqref{3=1} in a spatially inhomogeneous
half-space with $\rho =\rho (\mathbf{x})$, $c_{ijkl}=c_{ijkl}(\mathbf{x})$
may be written in Stroh-like format reminiscent of eqs.\ \eqref{=2} and %
\eqref{02-} as follows. Distinguish the coordinate normal to the surface, $%
y\equiv \mathbf{n}\cdot \mathbf{x}$, and the vector $\boldsymbol{z}\in 
\mathbb{R}^{2}$ representing the orthogonal two-dimensional space. The
governing equations then take the form 
\begin{equation}
\frac{\partial \boldsymbol{\mu }}{\partial y}=i\mathcal{N}\boldsymbol{\mu }
\label{=12}
\end{equation}%
with 
\begin{equation}
\begin{aligned} \boldsymbol{\mu } (y, \boldsymbol{z }) = \begin{pmatrix}
{\bf u} \\ i \mathbf{n} \boldsymbol{\sigma} \end{pmatrix} , \ \ &\mathcal{N}
= - \begin{pmatrix} \mathcal{T}^{-1}\mathcal{R}^+ & \mathcal{T}^{-1} \\
\mathcal{R}\mathcal{T}^{-1}\mathcal{R}^+ -\mathcal{Q}+\rho
\omega^2\mathbf{I} & \mathcal{R}\mathcal{T}^{-1} \end{pmatrix}, \\ &
\mathcal{T} = (n n),\ \ \mathcal{R} = ( -i\nabla_z ,\, n),\ \ \mathcal{Q} =
(-i\nabla_z ,\, -i\nabla_z ) , \end{aligned}  \label{=13}
\end{equation}%
where the definition of the 3$\times $3 matrix $\left( pq\right) =(p,q)$ has
been expanded to include the operator with components $%
(pq)_{jk}=p_{i}c_{ijkl}q_{l}$ for vector operators $\mathbf{p}$ and $\mathbf{%
q}$.

%Consider the situation where $\rho$,  $c_{ijkl}$ and hence
%$\mathcal{N}$ are ${\bf T}$-periodic in
%$\boldsymbol{z }$.

Let us consider the material with parameters that are $\mathbf{T}$-periodic
in $\boldsymbol{z}$: 
\begin{equation}
h\big(y,\boldsymbol{z}+n_{1}\mathbf{a}_{1}+n_{2}\mathbf{a}_{2}\big)=h(y,%
\boldsymbol{z})\quad \text{for }\ h=\rho ,c_{ijkl},  \label{19}
\end{equation}%
where $\boldsymbol{z}\in \mathbb{R}^{2}$, $n_{1},n_{2}\in \mathbb{Z}$, and
the linear independent translation vectors $\mathbf{a}_{1},\mathbf{a}_{2}\in 
\mathbb{R}^{2}$ define the irreducible unit cell $\mathbf{T}%
=\sum\nolimits_{j=1}^{2}t_{j}\mathbf{a}_{j}$ ($t_{j}\in \left[ 0,1\right] $)
of the periodic lattice. Let $\left\{ \mathbf{e}_{1}\mathbf{e}_{2}\right\} $
define an orthonormal base in $\mathbb{R}^{2}$ and denote 
\begin{equation}
\mathbf{a}_{j}=\mathbf{A e}_{j},\ \mathbf{b}_{j}=\left( \mathbf{A}%
^{-1}\right) ^{\mathrm{T}}\mathbf{e}_{j},\quad \Gamma =\{\mathbf{g}:\ 
\mathbf{g}=%
\begin{matrix}
\sum\nolimits_{j=1}^{2}%
\end{matrix}%
2\pi n_{j}\mathbf{b}_{j},\ (n_{1},n_{2})\in \mathbb{Z}^{2}\}.  \label{101}
\end{equation}%
The material parameters can therefore be expressed as 
\begin{equation}
h(y,\boldsymbol{z})=\sum\limits_{\mathbf{g}\in \Gamma }\mathrm{e}^{i\mathbf{g%
}\cdot \boldsymbol{z}}\,\hat{h}(y,\mathbf{g})\ \ \Leftrightarrow \ \ \hat{h}%
(y,\mathbf{g})=\frac{1}{|\mathbf{T}|}\int_{\mathbf{T}}\mathrm{d}\boldsymbol{z%
}\,\mathrm{e}^{-i\mathbf{g}\cdot \boldsymbol{z}}h(y,\boldsymbol{z})\quad 
\text{for }\ h={\mathbf{C}},\rho .  \label{212}
\end{equation}%
In practical terms the matrices are limited to finite size by restricting
the set of reciprocal lattice vectors to $\mathbf{g}\in \Gamma _{0}$ where $%
\Gamma _{0}\subset \Gamma $ comprises a finite number of elements, say $N$.

We look for solutions in the Floquet form 
\begin{equation}
\boldsymbol{\mu }(y,\boldsymbol{z})=\boldsymbol{\eta }(y,\boldsymbol{z})\,%
\mathrm{e}^{ik\mathbf{m}\cdot \boldsymbol{z}}  \label{=5}
\end{equation}%
for $\mathbf{T}$\textbf{-}periodic $\boldsymbol{\eta }(y,\boldsymbol{z})$.
The surface wavenumber vector $\mathbf{k}=k\mathbf{m}$ resides in the first
Brillouin zone of the reciprocal lattice $\Gamma $ and is otherwise
arbitrary. Employing a plane wave expansion 
\begin{equation}
\boldsymbol{\eta }(y,\boldsymbol{z})=\sum\nolimits_{\mathbf{g}\in \Gamma
_{0}}\mathrm{e}^{i\mathbf{g}\cdot \boldsymbol{z}}\,\widehat{\boldsymbol{\eta 
}}(y,\mathbf{g}),  \label{=6}
\end{equation}%
eq.\ \eqref{=12} becomes an ordinary differential equation for the $6N-$%
vector $\widetilde{\boldsymbol{\eta }}(y)$ comprised of all $\widehat{%
\boldsymbol{\eta }}(y,\mathbf{g})$, $\mathbf{g}\in \Gamma _{0}$, 
\begin{equation}
\frac{\mathrm{d}\widetilde{\boldsymbol{\eta }}}{\mathrm{d}y}=i\widetilde{%
\mathbf{N}}\widetilde{\boldsymbol{\eta }},\quad \widetilde{\mathbf{N}}=-%
\begin{pmatrix}
\widetilde{\mathbf{T}}^{-1}\widetilde{\mathbf{R}}^{+} & \widetilde{\mathbf{T}%
}^{-1} \\ 
\widetilde{\mathbf{R}}\widetilde{\mathbf{T}}^{-1}\widetilde{\mathbf{R}}^{+}-%
\widetilde{\mathbf{P}} & ~~\widetilde{\mathbf{R}}\widetilde{\mathbf{T}}^{-1}%
\end{pmatrix}%
.  \label{=7}
\end{equation}%
Here $\widetilde{\mathbf{P}}=\widetilde{\mathbf{P}}^{+}$, $\widetilde{%
\mathbf{T}}=\widetilde{\mathbf{T}}^{+}$ and $\widetilde{\mathbf{R}}$ are $%
3N\times 3N$ matrices with $3\times 3$ blocks associated with PWE
wavenumbers $\mathbf{g}$, $\mathbf{g}^{\prime }$, given by 
\begin{equation}
\begin{aligned} (\widetilde{ \text{P}} )_{{\bf g}{\bf g}'} &= (k+g,k+g')
({\bf g} - {\bf g}') - \omega^2 \hat{\rho}(y,{\bf g} - {\bf g}') \mathbf{I},
\\ (\widetilde{ \text{T}} )_{{\bf g}{\bf g}'} &= (n, n) ({\bf g} - {\bf
g}'), \ \ (\widetilde{ \text{R}} )_{{\bf g}{\bf g}'} = (k+g, n) ({\bf g} -
{\bf g}'), \end{aligned}  \label{-09}
\end{equation}%
where the definition of $(pq)_{jk}(\mathbf{g})$ is the natural extension of $%
(pq)_{jk}$ to include the Fourier transform, 
\begin{equation}
(pq)_{jk}(\mathbf{g})\equiv p_{i}\hat{c}_{ijkl}(y,\mathbf{g})q_{l}.
\label{=8}
\end{equation}%
Note that $\widetilde{\mathbf{T}}$ is negative definite and hence invertible.

Equation \eqref{=7} is valid regardless of whether the material properties
depend on $y$ or not. We consider here the case of a laterally periodic
material whose properties are periodic along the surface and uniform in the
depth direction, so that the system matrix $\widetilde{\mathbf{N}}$ is
constant. A method for treating the case of periodic $\widetilde{\mathbf{N}}%
(y)$ is discussed in \cite{Kutsenko12}. % and is briefly mentioned below.

\subsection{Surface impedance and related matrices via the matrix sign
function}

\label{5.2}

For constant $\widetilde{\mathbf{N}},$ the governing equation \eqref{=7} is
analogous to that in the uniform elastic half-space and its solution is in a
way analogous to (\ref{-91}), with two major differences. The first is that
we are now dealing with large, formally infinite, vectors and matrices (for
convenience, we continue to refer to $\widetilde{\mathbf{N}}$ of $6N\times
6N $ size). The second is that, in contrast to the real-valued Stroh matrix $%
\mathbf{N}$, the system matrix $\widetilde{\mathbf{N}}$ of \eqref{=7} is
generally complex. Its symmetry is 
\begin{equation}
\widetilde{\mathbf{N}}=\widetilde{\mathbf{K}}\widetilde{\mathbf{N}}^{+}%
\widetilde{\mathbf{K}},  \label{N+}
\end{equation}%
where $\widetilde{\mathbf{K}}$ is a $6N\times 6N$ equivalent of the $6\times
6$ matrix ${\mathbf{K}}$ with two zero and two identity blocks on and off
the diagonal. Denote the eigenvalues and eigenvectors of $i\widetilde{%
\mathbf{N}}$ by $\widetilde{\lambda }_{\alpha }$ and $\widetilde{\boldsymbol{%
\xi }}_{\alpha }$ As everywhere above, we restrict our attention to the
subsonic domain where, by virtue of (\ref{N+}), the set of $6N$ eigenvalues
of $i\widetilde{\mathbf{N}}$ can be split in two halves as 
\begin{equation}
\lambda _{\alpha }=-\lambda _{\alpha +3N}^{\ast }\ \ (\mathrm{Re}\lambda
_{\alpha }<0)\ \mathrm{for}\ \alpha =1,...,3N,  \label{883+}
\end{equation}%
which correspond to physical (decaying) and nonphysical (growing) modes.
Adopting the same partitioning for the eigenvectors, denote%
\begin{equation}
\widetilde{\mathbf{\Gamma }}=\left\Vert \widetilde{\boldsymbol{\xi }}_{1},...%
\widetilde{\boldsymbol{\xi }}_{3N},\widetilde{\boldsymbol{\xi }}_{3N+1}...%
\widetilde{\boldsymbol{\xi }}_{6N}\right\Vert = 
\begin{pmatrix}
\widetilde{\mathbf{A}}_{1} & \widetilde{\mathbf{A}}_{2} \\ 
\widetilde{\mathbf{L}}_{1} & \widetilde{\mathbf{L}}_{2}%
\end{pmatrix}%
.  \label{-4+}
\end{equation}%
It also follows from (\ref{N+}) that the orthogonality/completeness
relations in the subsonic domain hold in the form similar to (\ref{884}),
namely,%
\begin{equation}
\widetilde{\mathbf{\Gamma }}^{+}\widetilde{\mathbf{K}}\widetilde{\mathbf{%
\Gamma }}=\widetilde{\mathbf{K}}\quad \Leftrightarrow \quad \widetilde{%
\mathbf{\Gamma }}\mathbf{\widetilde{\mathbf{K}}\widetilde{\mathbf{\Gamma }}}%
^{+}=\widetilde{\mathbf{K}}.  \label{884+}
\end{equation}%
Note that, in contrast to the uniform case (\ref{883}), $\widetilde{\mathbf{%
\xi }}_{\alpha }$ and $\widetilde{\boldsymbol{\xi }}_{\alpha +3N}$ are not
complex conjugated, i.e. $\widetilde{\mathbf{\Gamma }}^{\ast }\neq 
\widetilde{\mathbf{\Gamma }}\mathbf{\widetilde{\mathbf{K}}}$ and so (\ref%
{884+}) has no equivalent that would be a generalization of (\ref{883-1}).
This is reminiscent of the Stroh formalism in cylindrical coordinates, see 
\cite{Norris10} and \cite[\S 3d]{Shuvalov03}.

Now we can follow the line of derivation proposed in \S \ref{3.2}. Introduce
the sign function associated with $i\widetilde{\mathbf{N}},$ namely,%
\begin{equation}
(\sign i\widetilde{\mathbf{N}})\widetilde{\boldsymbol{\xi }}_{\alpha }=-%
\widetilde{\boldsymbol{\xi }}_{\alpha },\ \ (\sign i\widetilde{\mathbf{N%
}})\widetilde{\boldsymbol{\xi }}_{\alpha +3}=\widetilde{\boldsymbol{\xi }}%
_{\alpha +3},\ \ \alpha =1,...,3N  \label{eigen+}
\end{equation}%
and denote its blocks as 
\begin{equation}
\sign i\widetilde{\mathbf{N}}=i%
\begin{pmatrix}
\widetilde{\mathbf{S}} & \widetilde{\mathbf{Q}} \\ 
\widetilde{\mathbf{B}} & \widetilde{\mathbf{S}}^{+}%
\end{pmatrix}%
\ \mathrm{with}\ \widetilde{\mathbf{Q}}=\widetilde{\mathbf{Q}}^{+},\ 
\widetilde{\mathbf{B}}=\widetilde{\mathbf{B}}^{+}, \ \text{tr}\, \widetilde{%
\mathbf{S}}=0.  \label{def+}
\end{equation}%
By definition (\ref{eigen+}), $(\sign i\widetilde{\mathbf{N}})^{2}=%
\mathbf{I}$ and so%
\begin{equation}
\widetilde{\mathbf{B}}\widetilde{\mathbf{S}}+\widetilde{\mathbf{S}}^{+}%
\widetilde{\mathbf{B}}=\mathbf{0},\ \ \widetilde{\mathbf{S}}\widetilde{%
\mathbf{Q}}+\widetilde{\mathbf{Q}}\mathbf{\widetilde{\mathbf{S}}}^{+}=%
\mathbf{0},\ \ \mathbf{\widetilde{\mathbf{S}}}^{2}+\widetilde{\mathbf{Q}}%
\widetilde{\mathbf{B}}=-\mathbf{I.}  \label{7.16+}
\end{equation}%
Using (\ref{884+}), the spectral decomposition is (cf.\ eq.\ \eqref{spectral}%
) 
\begin{equation}
\sign i\widetilde{\mathbf{N}}=\widetilde{\mathbf{\Gamma }}\mathrm{diag}%
(-\mathbf{I},\mathbf{I})\widetilde{\mathbf{K}}\mathbf{\widetilde{\mathbf{%
\Gamma }}}^{+}\widetilde{\mathbf{K}}= 
\begin{pmatrix}
-\mathbf{I} & \mathbf{0} \\ 
\mathbf{0} & -\mathbf{I}%
\end{pmatrix}
+2%
\begin{pmatrix}
\widetilde{\mathbf{A}}_{2}\widetilde{\mathbf{L}}_{1}^{+} & \widetilde{%
\mathbf{A}}_{2}\widetilde{\mathbf{A}}_{1}^{+} \\ 
\widetilde{\mathbf{L}}_{2}\widetilde{\mathbf{L}}_{1}^{+} & \widetilde{%
\mathbf{L}}_{2}\widetilde{\mathbf{A}}_{1}^{+}%
\end{pmatrix}%
.  \label{spectral+}
\end{equation}%
Hence from (\ref{def+}) and (\ref{spectral+})%
\begin{equation}
\widetilde{\mathbf{S}}=i(\mathbf{I} - 2\widetilde{\mathbf{A}}_{2}\widetilde{%
\mathbf{L}}_{1}^{+} ),\quad \widetilde{\mathbf{Q}}=- 2i\widetilde{\mathbf{A}}%
_{2}\widetilde{\mathbf{A}}_{1}^{+},\quad \widetilde{\mathbf{B}}=-2i%
\widetilde{\mathbf{L}}_{2}\widetilde{\mathbf{L}}_{1}^{+}.  \label{==3+}
\end{equation}

Assuming invertibility of $\widetilde{\mathbf{Q}}$ and hence of $\widetilde{%
\mathbf{A}}_{1}$ and $\widetilde{\mathbf{A}}_{2}$, introduce the surface
impedance $\widetilde{\mathbf{Z}}=i\widetilde{\mathbf{L}}_{1}\widetilde{%
\mathbf{A}}_{1}^{-1}.$ From (\ref{884+}) $\widetilde{\mathbf{Z}}$ is
Hermitian and combining (\ref{eigen+})$_{1}$ with (\ref{def+}) relates $%
\widetilde{\mathbf{Z}}$ to the blocks of $\sign i\widetilde{\mathbf{N}}$%
. Thus%
\begin{equation}
\widetilde{\mathbf{Z}}=i\widetilde{\mathbf{L}}_{1}\widetilde{\mathbf{A}}%
_{1}^{-1}(=\widetilde{\mathbf{Z}}^{+})\ \Leftrightarrow \ \widetilde{\mathbf{%
Z}}=-\widetilde{\mathbf{Q}}^{-1}(\mathbf{I}+i\widetilde{\mathbf{S}}).
\label{7.22+}
\end{equation}%
Note from (\ref{7.22+}) that $\det \big(\mathbf{I}+i\widetilde{\mathbf{S}}%
\big)$ is real (since $\widetilde{\mathbf{Z}}$\ and $\widetilde{\mathbf{Q}}$
are Hermitian) and, unlike the case of a uniform medium (hence of real $%
\mathbf{Q}$ and $\mathbf{S}$), the determinant of $ \widetilde{\mathbf{S}}$ 
is generally nonzero. 
%on account of \eqref{7.16+}$_{2}$ and the assumed invertibility of $\widetilde{\mathbf{Q}}$%
.

Finally, the evaluation of $\sign i\widetilde{\mathbf{N}}$ is possible
by different methods, including the integral representation analogous to %
\eqref{82}, yielding 
\begin{align}  \label{333}
\widetilde{\mathbf{S}} &= - \big\langle \widetilde{\mathbf{T}}_\theta^{-1} 
\widetilde{\mathbf{R}}_\theta^+ \big\rangle , \ \ \widetilde{\mathbf{Q}}= - %
\big\langle \widetilde{\mathbf{T}}_\theta^{-1} \big\rangle , \ \ \widetilde{%
\mathbf{B}} = \big\langle \widetilde{\mathbf{T}}_{\theta+\frac{\pi}2} - 
\widetilde{\mathbf{R}}_\theta\widetilde{\mathbf{T}}_\theta^{-1} \widetilde{%
\mathbf{R}}_\theta^+ \big\rangle , \\
& \text{where} \ \ \left\{ \begin{aligned} \widetilde{\mathbf{T}}_{\theta
}=\cos ^{2}\theta \widetilde{\mathbf{T}}+\sin ^{2}\theta
\widetilde{\mathbf{P}}-\sin \theta \cos \theta
(\widetilde{\mathbf{R}}+\widetilde{\mathbf{R}}^{+}), \\
\widetilde{\mathbf{R}}_{\theta }=\cos ^{2}\theta \widetilde{\mathbf{R}}+\sin
^{2}\theta \widetilde{\mathbf{R}}^{+}+\sin \theta \cos \theta
(\widetilde{\mathbf{T}}+\widetilde{\mathbf{P}}).\end{aligned} \right.  \notag
\end{align}

\subsubsection{Dispersion equation}

Existence of a surface wave with a speed $v=v_{s}$ under the traction-free
boundary implies that there must exist some non-zero vector $\widetilde{%
\mathbf{v}}$ at $v_{s}$ such that 
\begin{equation}
\sign i\widetilde{\mathbf{N}}%
\begin{pmatrix}
\widetilde{\mathbf{v}} \\ 
\widetilde{\mathbf{0}}%
\end{pmatrix}%
=%
\begin{pmatrix}
-\widetilde{\mathbf{v}} \\ 
\widetilde{\mathbf{0}}%
\end{pmatrix}%
\quad \Rightarrow \quad \left\{ \begin{aligned} (\widetilde{\mathbf I} +i
\widetilde{\mathbf S} ) \widetilde{\bf v} &= 0, \\ \widetilde{\mathbf
B}\widetilde{\bf v} &=0, \end{aligned}\right. \ \ \Leftrightarrow \ \
\left\{ \begin{aligned} \widetilde{\mathbf A}_2 \widetilde{\mathbf L}_1^+
\widetilde{\bf v} &= 0, \\ \widetilde{\mathbf L}_2 \widetilde{\mathbf L}_1^+
\widetilde{\bf v} &=0, \end{aligned}\right.  \label{-37}
\end{equation}%
where (\ref{def+}) and \eqref{==3+} have been used. The situation is
reminiscent of eq.\ (\ref{-0}) for a uniform half-space, certainly apart
from the fact that eq.\ (\ref{-37}) involves matrices of a large, formally
infinite, size. Another particularity is that the equality $\det \widetilde{%
\mathbf{B}}=0$ may signify occurrence of either a physical (if $\det 
\widetilde{\mathbf{L}}_{1}=0$) or a nonphysical (if $\det \widetilde{\mathbf{L}%
}_{2}=0$) surface wave solution. At the same time, the condition $\det (%
\mathbf{I+}i\widetilde{\mathbf{S}})~(=\det 2\widetilde{\mathbf{A}}_{2}%
\widetilde{\mathbf{L}}_{1}^{+})=0$ is both necessary and sufficient for the
physical wave in, specifically, the subsonic domain, where $\widetilde{%
\mathbf{Q}}$ and $\widetilde{\mathbf{A}}_{1},$ $\widetilde{\mathbf{A}}_{2}$
are invertible (and so $\widetilde{\mathbf{Z}}$ exists) as can be argued
similarly as in \S 3.3. Interestingly, this is no longer so in the upper
overlapping band gaps of the Floquet spectrum, where the condition \eqref{-37} is recast
in the form with a positive definite matrix 
\begin{equation}
\big((\widetilde{\mathbf{I}}+i\widetilde{\mathbf{S}})^{+}(\widetilde{\mathbf{%
I}}+i\widetilde{\mathbf{S}})+\widetilde{\mathbf{B}}^{+}\widetilde{\mathbf{B}}%
\big)\widetilde{\mathbf{v}}=0\ \ \ (\Leftrightarrow \ \ \widetilde{\mathbf{L}%
}_{1}\big(\widetilde{\mathbf{A}}_{2}^{+}\widetilde{\mathbf{A}}_{2}+%
\widetilde{\mathbf{L}}_{2}^{+}\widetilde{\mathbf{L}}_{2}\big)\widetilde{%
\mathbf{L}}_{1}^{+}\widetilde{\mathbf{v}}=0),  \label{-29}
\end{equation}%
which provides a single dispersion equation%
\begin{equation}
\det \big((\widetilde{\mathbf{I}}+i\widetilde{\mathbf{S}})^{+}(\widetilde{%
\mathbf{I}}+i\widetilde{\mathbf{S}})+\widetilde{\mathbf{B}}^{+}\widetilde{%
\mathbf{B}}\big)=0\ \ (\Leftrightarrow \ \ |\det \widetilde{\mathbf{L}}%
_{1}|^{2}=0).  \label{+29}
\end{equation}%
Note that those considerations above which used spectral decomposition under
the assumption of distinct eigenvalues can be reproduced in the invariant
terms of the projector matrix, see \cite{Kutsenko12}.

Significant simplification follows for the case of a symmetric unit cell
where the Fourier expansion \eqref{212}$_{1}$ reduces to a cosine expansion
with real valued coefficients. In consequence the matrix $\widetilde{\mathbf{%
N}}$ is real and hence the basic conclusions obtained above for the case of
a uniform half-space can be extended to the present case in hand. In
particular by analogy with eq.\ (\ref{36}) it follows that the dispersion
equation on the subsonic surface wave speed $v=v_{s}$ can be taken in any of
the following equivalent real-valued forms 
\begin{equation*}
\det \widetilde{\mathbf{Z}}=0\ \Leftrightarrow \ \det (\widetilde{\mathbf{I}}%
+i\widetilde{\mathbf{S}})=0\Leftrightarrow \ \det \widetilde{\mathbf{B}}=0\
(\Rightarrow \ \mathrm{tr}\,\text{adj}\widetilde{\mathbf{B}}=0),
\end{equation*}%
where {$\text{adj}\widetilde{\mathbf{B}}$ is the adjugate matrix.} Note that
the double zero of $\det \mathbf{B}$\ at $v=v_{s}$ makes the two other forms
of the dispersion equation more appealing for numerical evaluation of $v_{s}$%
.

\subsubsection{Examples}

Numerical results are presented for a laterally periodic half-space composed
of layers of two alternating materials. Interfaces between the layers are
normal to the surface of the half-space. Note that a periodically bilayered
structure infinite along the periodicity axis can be seen as a periodically
tri-layered with a symmetric unit cell and hence with a real matrix $%
\widetilde{\mathbf{N}},$ see \S 5.2.1. We assume bimaterial structures of
equidistant layers of any two of the three isotropic solids: copper (Cu,
Young's modulus $E$ = 115 GPa, Poisson's ratio $\nu $ = 0.355, $\rho $ =
8920 kg/m$^{3}$), aluminum (Al, 69 GPa, 0.334, 2700 kg/m$^{3}$) or steel
(St, 203 GPa, 0.29, 7850 kg/m$^{3}$). The surface wave propagation direction 
$\mathbf{m,}$ that is the direction of the wavenumber vector $\mathbf{k}=k%
\mathbf{m}$ (see (\ref{=5})), is measured by the angle $\psi $ between $%
\mathbf{m}$ and the plane of interlayer interfaces, so that $\psi =0$
corresponds to $\mathbf{m}$ in the layering direction, i.e., normal to the
interfaces. The values of wavenumber $k$ considered are restricted to the
first Brillouin zone defined by the cell length in the layering direction,
which is taken as unity.

\begin{figure}[tbp]
\centering
 \subfigure{
  \includegraphics[width=4.in]{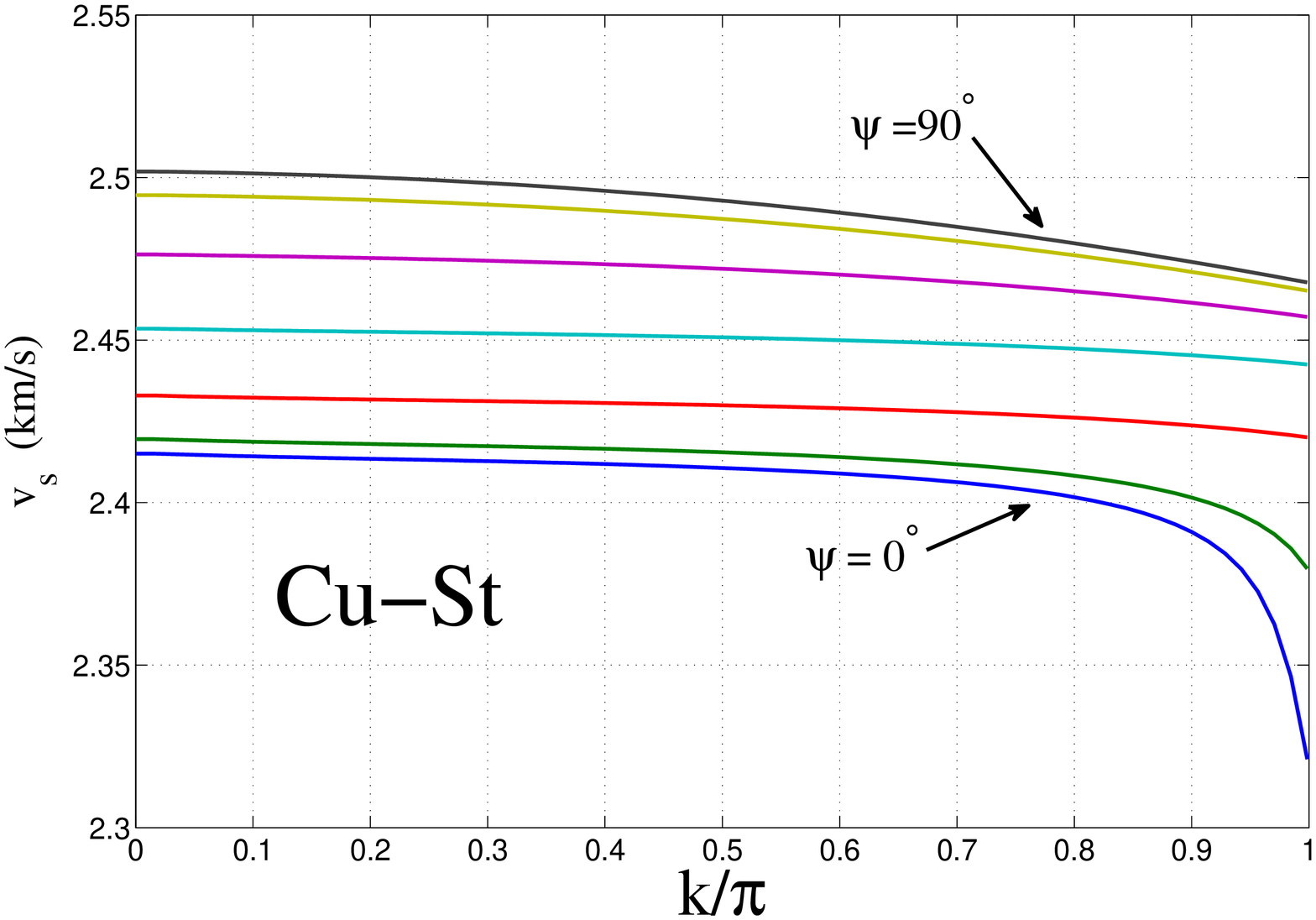}
    }
 \subfigure{  \includegraphics[width=4.in]{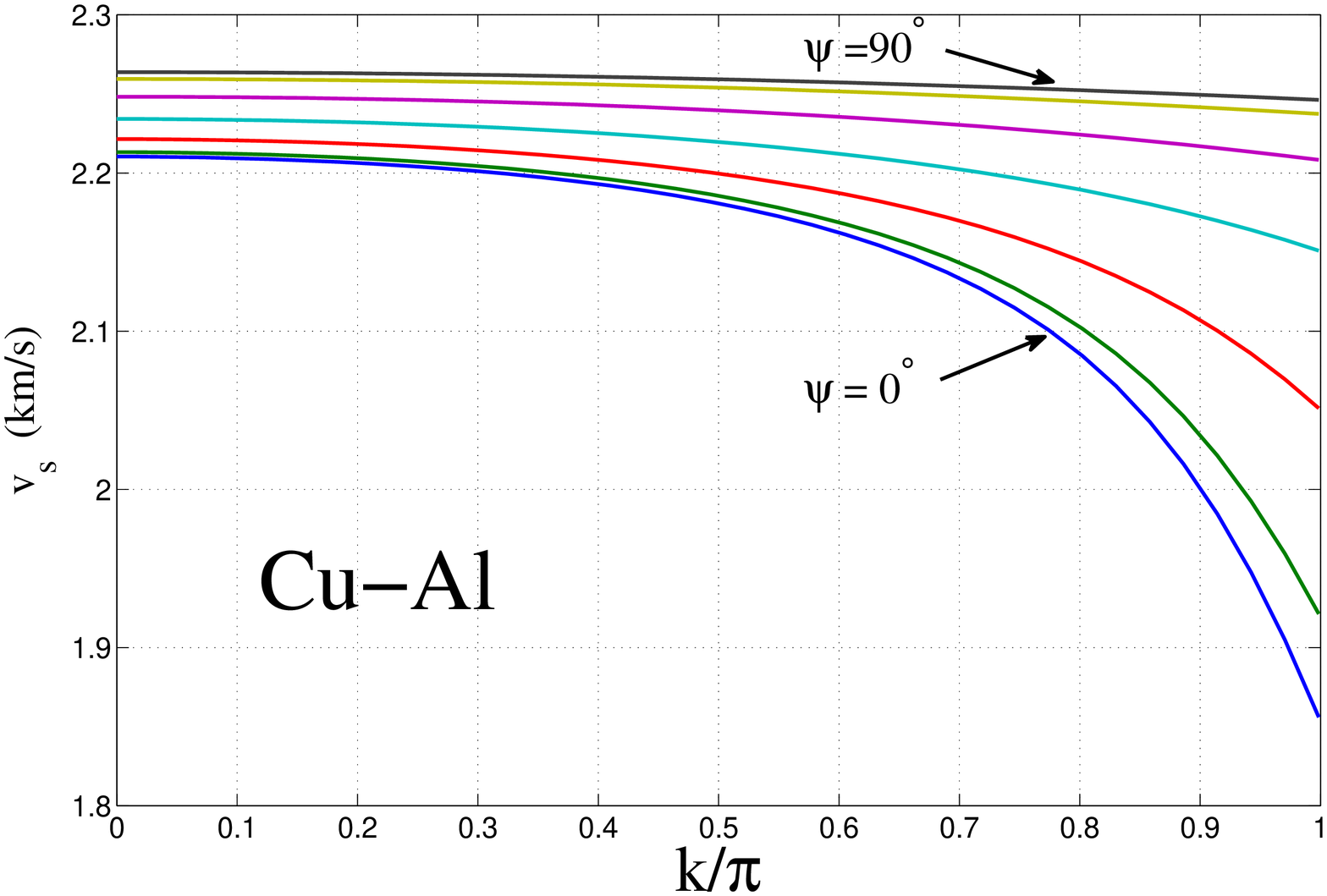}
   }
 \subfigure{
 \includegraphics[width=4.in]{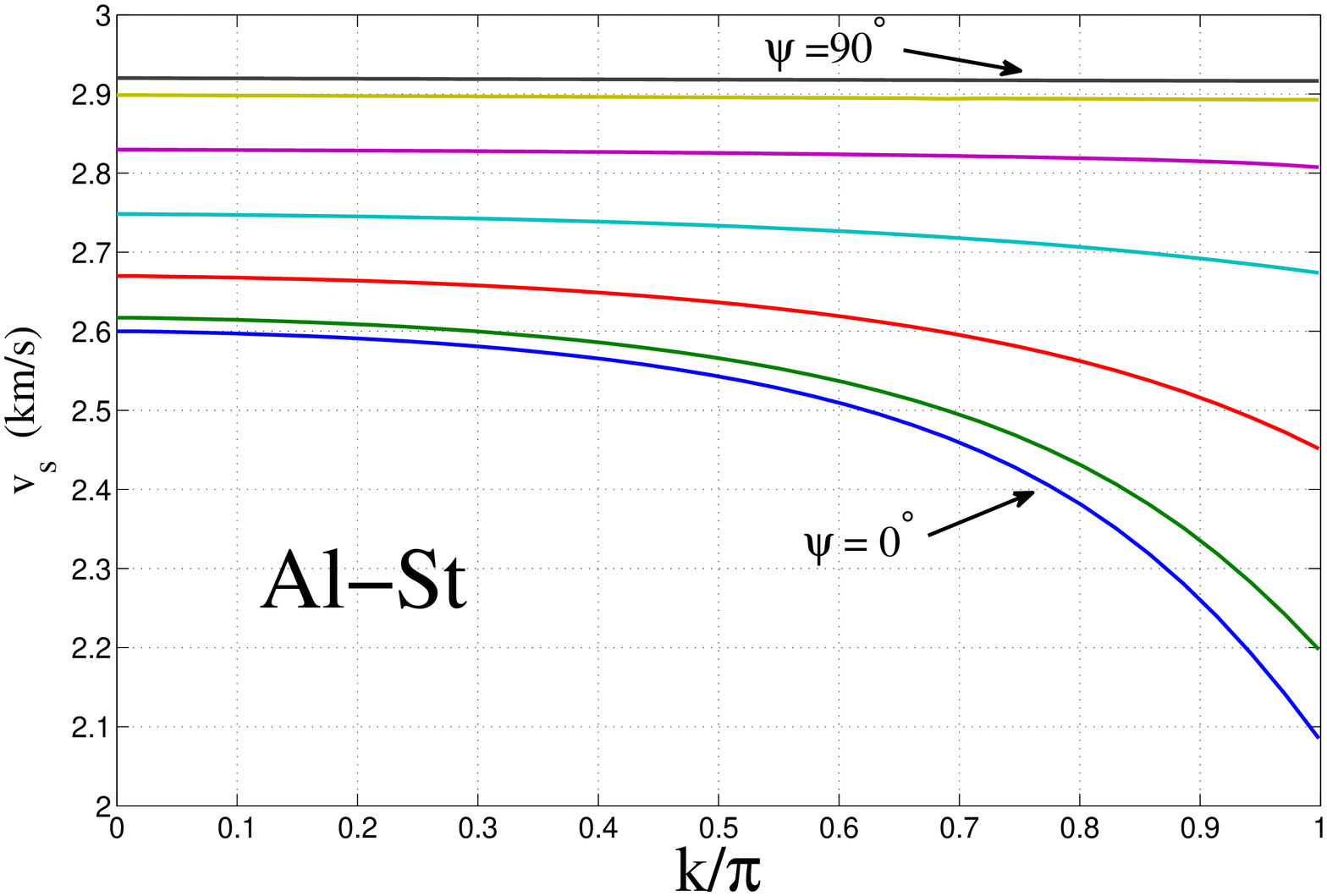}
   }
\caption{The surface wave speed $v_s$ for three laterally periodic bimaterials
made of equal volume fractions of any two of copper, steel or aluminum (Cu,
St, Al). The propagation direction $\mathbf{m}$ varies from $\protect\psi =0$
(normal to the layers) to $\protect\psi =90^{\circ }$ in increments of $%
15^{\circ }$. }
\label{fig1}
\end{figure}

Figure \ref{fig1} demonstrates the computed surface wave speed $v_{s}$ as a
function of wave number $k$ for seven distinct directions of $\mathbf{m}$.
The curves displayed are azimuthal cross-sections of the subsonic dispersion
surface $v_{s}=\omega (k\mathbf{m})/k=v_{s}(k,\psi )$. The numerical results
show that surface wave speed decreases as a function of $k$ and increases
with $\psi $ for all  bimaterials considered. The dependence of $v_{s}$
on $k$ at fixed $\psi $ is greatest for waves traveling across the
interfaces $(\psi =0)$ and least for waves traveling along the interfaces $%
(\psi =90^{\circ })$.

\section{Conclusions}

Use of the matrix sign function provides a new and broader perspective of
the surface wave problem in anisotropic media. It straightens out the
methodology of the underlying matrix formalism and offers a direct method to
compute the matrices involved. Starting from the defining property of the
matrix sign function \eqref{274} we have obtained known relations for the
Barnett-Lothe matrices, the impedance matrix and the dispersion equation for
the surface wave speed, see eqs.\ \eqref{7.16}, \eqref{7.22} and \eqref{36}.
These expressions are achieved using only the sign function of  $i$ times the
Stroh matrix $\mathbf{N}=\mathbf{K}\mathbf{N}^{T}\mathbf{K}$ without
specifying its method of evaluation. An integral representation \eqref{-55-}
for the matrix sign function, combined with the explicit structure of the
Stroh matrix \eqref{=22}, leads immediately to the Barnett-Lothe integral
relations \eqref{82}. In this paper we have concentrated on using the matrix
sign function for direct formulation of the dispersion equation, without
discussing further properties of the Barnett-Lothe matrices which underlie
the existence and uniqueness considerations in the surface wave theory \cite%
{Barnett73,Lothe76,Chadwick77}. We have constructed an explicit solution of
the matrix Riccati equation for the impedance $\mathbf{Z}$ by rewriting the
Riccati equation in a form \eqref{-84} that involves the matrix sign
function. Apart from providing for the first time a direct solution of the
Riccati equation, the use of the matrix sign function shows how this
nonlinear equation is intimately related with the Stroh matrix.

Perhaps the greatest advantage gained by using the matrix sign function is
that it provides a natural formalism for framing the problem of subsonic
surface waves in laterally periodic half-spaces which are inhomogeneous
along the surface and uniform in the depth direction. We have shown how much
of the structure for the homogeneous case carries over to the case of
laterally periodic materials. For instance, the conditions for surface waves %
\eqref{-37} and the form of the block matrices in the matrix sign function %
\eqref{333} mirror their counterparts for the homogeneous case, eqs.\ %
\eqref{-0} and \eqref{82}, respectively. Naturally, there are major
differences between the problems. Conditions for the existence of surface
waves in periodically inhomogeneous materials have only been recently
established \cite{Liu11} but the methods that have been proposed for finding
them are not as straightforward as for the homogeneous case. The approach
that we have presented for the laterally periodic case, being linked via the
matrix sign function to the well known formalism for the homogeneous
half-space, offers, we believe, a clear and logical route for finding
surface waves. Future work will examine this calculation method and the properties of the
subsonic  waves in more detail and will also consider supersonic
solutions in the upper stopbands and passbands - the extension motivated by
the classical paper by Alshits and Lothe \cite{Alshits81}. 
%[V.I. Alshits, J. Lothe. Comments on the relation between surface wave theory and the theory of reflection. Wave Motion, 1981, V.3, 297-310.]

%%%%%%%%%%%%%%%%%%%%%%%%%
 
\section*{Appendix: Sign function and related matrix functions}
 
\label{appendix}

The matrix sign function of $\mathbf{M}$ is closely related to other
standard matrix functions. The matrix \textit{projector} functions $\mathbf{P%
}_+$ and $\mathbf{P}_-$ are defined 
\begin{equation}  \label{33}
\mathbf{P}_\pm \boldsymbol{\xi } = 
\begin{cases}
\boldsymbol{\xi }, & \boldsymbol{\xi } \in \mathcal{L}_\pm , \\ 
\mathbf{0}, & \boldsymbol{\xi } \in \mathcal{L}_\mp ,%
\end{cases}
\ \ \text{where} \ \mathcal{L}_\pm = \big\{ \boldsymbol{\xi }: \, \mathbf{M}%
\boldsymbol{\xi }= \lambda \boldsymbol{\xi }, \ \mathrm{Re} \,\lambda \, 
{\tiny 
\begin{matrix}
> \\ 
<%
\end{matrix}
} \, 0 , \big\},
\end{equation}
and therefore 
\begin{equation}  \label{-7}
\sign \mathbf{M} = \mathbf{P}_+ - \mathbf{P}_- , \ \ \mathbf{I} = 
\mathbf{P}_+ + \mathbf{P}_- \quad \Leftrightarrow \quad \mathbf{P}_\pm =
\frac 12 \mathbf{I} \pm \frac12 \sign \mathbf{M} .
\end{equation}
The projector functions can be expressed via integration, e.g.\ \cite{Bai98} 
\begin{equation}  \label{1}
\mathbf{P}_\pm = \frac 1{2\pi i}\int_{C_\pm} \mathrm{d} z\, (z \mathbf{I}
\mp \mathbf{M})^{-1}
\end{equation}
where $C_\pm$ counter-clockwise encloses the finite right(left)-half plane $%
\{z:\, \mathrm{Re}\, (\pm z)> 0, \ |z|<R\}$, $R$ arbitrarily large. The
Schwartz-Christoffel transformation $z\to (z-1)(z+1)^{-1}$ reduces this to
an integral around the unit circle 
\begin{equation}  \label{2}
\mathbf{P}_\pm = \frac 1{2\pi i}\int\limits_{|z|=1-0} \mathrm{d} z\, \big(z 
\mathbf{I} - \mathbf{F}^{\pm 1} \big)^{-1}, \quad \mathbf{F} = (\mathbf{M} - 
\mathbf{I})(\mathbf{M} + \mathbf{I})^{-1} .
\end{equation}
While $\mathbf{F} ^{\pm 1}$ are singular if $\mathbf{M} $ possesses
eigenvalues $\mp 1$, we note that \eqref{1} is unchanged for $\mathbf{M}\to a%
\mathbf{M} +i b\mathbf{I} $, $a,\, b\in \mathbb{R}$, $a>0$, and \eqref{2} is
invariant for $\mathbf{F}\to (\mathbf{M} - \alpha\mathbf{I})(\mathbf{M} +
\alpha^*\mathbf{I})^{-1} $, $\alpha \in \mathbb{C}$, $\mathrm{Re} \, \alpha
>0$, and hence \eqref{2} can always be made regular by selecting $\alpha $
in the right half-plane. The integral representation \eqref{-55-} for the
matrix sign function may be easily obtained from eqs.\ \eqref{-7} and %
\eqref{2}.

The \textit{disk} function \cite{Higham:2008:FM} is defined such that if $%
\mathbf{M} = \mathbf{U} ${\scriptsize {$%
\begin{pmatrix}
\mathbf{J}_1 & \mathbf{0} \\ 
\mathbf{0} & \mathbf{J}_2%
\end{pmatrix}%
$}} $\mathbf{U}^{-1}$ where the eigenvalues of $\mathbf{J}_1$, $\mathbf{J}_2$
have magnitudes $<1$, $>1$, respectively, then $\disk \mathbf{M} = 
\mathbf{U} ${\scriptsize {$%
\begin{pmatrix}
\mathbf{I} & \mathbf{0} \\ 
\mathbf{0} & \mathbf{0}%
\end{pmatrix}%
$}}$\mathbf{U}^{-1}$. It follows that 
\begin{equation}  \label{0-8}
\sign\mathbf{M} = \mathbf{I} - 2 \disk \big( ( \mathbf{M} -%
\mathbf{I} )^{-1} ( \mathbf{M} +\mathbf{I} ) \big).
\end{equation}

%%%%%%%%%%%%%%%%%%%%%%%%%%%%%%%%%%%%%%%%%%%%%%%%%%%%%%%%%%%%%%%%%%%%%%%%%%

\subsection*{Acknowledgment}

A.N.N. acknowledges support from Le Conseil Régional d’Aquitaine and the program CADMO of the cluster Advanced Materials in Aquitaine (GIS-AMA). A.A.K. acknowledges support from Mairie
de Bordeaux.

%\bibliography{../../SHARED_BIBLIOGRAPHY/AN_BIG_BIB}

\begin{thebibliography}{32}
\providecommand{\natexlab}[1]{#1}
\providecommand{\url}[1]{\texttt{#1}}
\expandafter\ifx\csname urlstyle\endcsname\relax
  \providecommand{\doi}[1]{doi: #1}\else
  \providecommand{\doi}{doi: \begingroup \urlstyle{rm}\Url}\fi

\bibitem[Stroh(1962)]{Stroh62}
A.~N. Stroh.
\newblock Steady state problems in anisotropic elasticity.
\newblock \emph{J. Math. Phys.}, 41:\penalty0 77–--103, 1962.

\bibitem[Barnett and Lothe(1973)]{Barnett73}
D.~M. Barnett and J.~Lothe.
\newblock Synthesis of the sextic and the integral formalism for dislocations,
  {G}reens functions, and surface waves in anisotropic elastic solids.
\newblock \emph{Phys. Norv.}, 7:\penalty0 13--19, 1973.

\bibitem[Lothe and Barnett(1976)]{Lothe76}
J.~Lothe and D.~M. Barnett.
\newblock On the existence of surface-wave solutions for anisotropic elastic
  half-spaces with free surface.
\newblock \emph{J. Appl. Phys.}, 47\penalty0 (2):\penalty0 428--433, 1976.

\bibitem[Chadwick and Smith(1977)]{Chadwick77}
P.~Chadwick and G.~D. Smith.
\newblock {Foundations of the theory of surface waves in anisotropic elastic
  materials}.
\newblock \emph{Adv. Appl. Mech.}, 17:\penalty0 303--376, 1977.

\bibitem[Ting(1996)]{Ting96}
T.~C.~T. Ting.
\newblock \emph{Anisotropic Elasticity: Theory and Applications}.
\newblock Oxford University Press, 1996.

\bibitem[Barnett(2000)]{Barnett00}
D.~M. Barnett.
\newblock {Bulk, surface, and interfacial waves in anisotropic linear elastic
  solids}.
\newblock \emph{Int. J. Solids Struct.}, 37\penalty0 (1-2):\penalty0 45--54,
  2000.
\newblock \doi{10.1016/S0020-7683(99)00076-1}.

\bibitem[Norris and Shuvalov(2010)]{Norris10}
A.~N. Norris and A.~L. Shuvalov.
\newblock Wave impedance matrices for cylindrically anisotropic radially
  inhomogeneous elastic materials.
\newblock \emph{Q. J. Mech. Appl. Math.}, 63:\penalty0 1--35, 2010.

\bibitem[Biryukov et~al.(1995)Biryukov, Gulyaev, Krylov, and
  Plessky]{Biryukov95}
V.~Biryukov, Yu.~V. Gulyaev, V.~V. Krylov, and V.~P. Plessky.
\newblock \emph{Surface Acoustic Waves in Inhomogeneous Media}.
\newblock Springer, Berlin, 1995.

\bibitem[Barnett and Lothe(1985)]{Barnett85}
D.~M. Barnett and J.~Lothe.
\newblock Free surface ({R}ayleigh) waves in anisotropic elastic half-spaces:
  The surface impedance method.
\newblock \emph{Proc. R. Soc. A}, 402\penalty0 (1822):\penalty0 135--152, 1985.
\newblock \doi{10.2307/2397800}.

\bibitem[Higham(1994)]{Higham:1994:MSD}
N.~J. Higham.
\newblock The matrix sign decomposition and its relation to the polar
  decomposition.
\newblock \emph{Linear Algebra Appl.}, 212/213:\penalty0 3--20, 1994.

\bibitem[Higham(2008)]{Higham:2008:FM}
N.~J. Higham.
\newblock \emph{Functions of Matrices: {Theory} and Computation}.
\newblock SIAM, Philadelphia, PA, 2008.

\bibitem[Higham()]{Higham}
N.~J. Higham.
\newblock {The Matrix Computation Toolbox}.
\newblock \url{http://www.ma.man.ac.uk/~higham/mctoolbox}.

\bibitem[Kenney and Laub(1995)]{Kenney:1995:MSF}
C.~S. Kenney and A.~J. Laub.
\newblock The matrix sign function.
\newblock \emph{IEEE Trans. Automat. Control}, 40\penalty0 (8):\penalty0
  1330--1348, 1995.

\bibitem[Roberts(1980)]{Roberts:1980:LMR}
J.~D. Roberts.
\newblock Linear model reduction and solution of the algebraic {Riccati}
  equation by use of the sign function.
\newblock \emph{Int. J. Control}, 32\penalty0 (4):\penalty0 677--687, 1980.

\bibitem[Gundersen and Lothe(1987)]{Gundersen87}
S.~A. Gundersen and J.~Lothe.
\newblock A new method for numerical calculations in anisotropic elasticity
  problemss.
\newblock \emph{Phys. Stat. Sol. (B)}, 143\penalty0 (1):\penalty0 73--85, 1987.
\newblock \doi{10.1002/pssb.2221430108}.

\bibitem[Condat and Kirchner(1987)]{Condat87}
M.~Condat and H.~O.~K. Kirchner.
\newblock Computational anisotropic elasticity.
\newblock \emph{Phys. Stat. Sol. (B)}, 144\penalty0 (1):\penalty0 137--143,
  1987.
\newblock \doi{10.1002/pssb.2221440112}.

\bibitem[Koc et~al.(1994)Koc, Bakkaloglu, and Shieh]{Koc94}
C.~K. Koc, B.~Bakkaloglu, and L.~S. Shieh.
\newblock {Computation of the matrix sign function using continued fraction
  expansion}.
\newblock \emph{IEEE Trans. Automatic Control}, 39\penalty0 (8):\penalty0
  1644--1647, 1994.
\newblock \doi{10.1109/9.310041}.

\bibitem[Ingebrigtsen and Tonning(1969)]{Ingebrigtsen69}
K.~A. Ingebrigtsen and A.~Tonning.
\newblock Elastic surface waves in crystals.
\newblock \emph{Phys. Rev.}, 184\penalty0 (3):\penalty0 942--951, 1969.
\newblock \doi{10.1103/PhysRev.184.942}.

\bibitem[Honein et~al.(1991)Honein, Braga, Barbone, and Herrmann]{Honein91}
B.~Honein, A.~M.~B. Braga, P.~Barbone, and G.~Herrmann.
\newblock Wave propagation in piezoelectric layered media with some
  applications.
\newblock \emph{J. Intell. Mater. Sys. Struct.}, 2\penalty0 (4):\penalty0
  542--557, 1991.
\newblock \doi{10.1177/1045389X9100200408}.

\bibitem[Wang and Rokhlin(2002)]{Wang02}
L.~Wang and S.~I. Rokhlin.
\newblock Recursive impedance matrix method for wave propagation in stratified
  media.
\newblock \emph{Bull. Seism. Soc. Am.}, 92:\penalty0 1129--1135, 2002.

\bibitem[Shuvalov and Every(2002)]{Shuvalov02}
A.~V. Shuvalov and A.~G. Every.
\newblock Some properties of surface acoustic waves in anisotropic-coated
  solids, studied by the impedance method.
\newblock \emph{Wave Motion}, 36:\penalty0 257--253, 2002.
\newblock \doi{10.1016/S0165-2125(02)00013-6}.

\bibitem[Hosten and Castaings(2003)]{Hosten03}
B.~Hosten and M.~Castaings.
\newblock Surface impedance matrices to model the propagation in multilayered
  media.
\newblock \emph{Ultrasonics}, 41\penalty0 (7):\penalty0 501--507, 2003.
\newblock \doi{10.1016/S0041-624X(03)00167-7}.

\bibitem[Fu(2003)]{Fu03}
Y.~B. Fu.
\newblock Existence and uniqueness of edge waves in a generally anisotropic
  elastic plate.
\newblock \emph{Q. J. Mech. Appl. Math.}, 56\penalty0 (4):\penalty0 605--616,
  2003.
\newblock \doi{10.1093/qjmam/56.4.605}.

\bibitem[Fu and Kaplunov(2012)]{Fu12}
Y.~B. Fu and J.~Kaplunov.
\newblock {Analysis of localized edge vibrations of cylindrical shells using
  the Stroh formalism}.
\newblock \emph{Math. Mech. Solids}, 17\penalty0 (1):\penalty0 59--66, 2012.
\newblock \doi{10.1177/1081286511412442}.

\bibitem[Biryukov(1985)]{Biryukov85}
S.~V. Biryukov.
\newblock Impedance method in the theory of elastic surface waves.
\newblock \emph{Sov. Phys. Acoust.}, 31:\penalty0 350--354, 1985.

\bibitem[Caviglia and Morro(2002)]{Caviglia02}
G.~Caviglia and A.~Morro.
\newblock Wave reflection and transmission from anisotropic layers through
  {R}iccati equations.
\newblock \emph{Q. J. Mech. Appl. Math.}, 55:\penalty0 93--107, 2002.

\bibitem[Fu and Mielke(2002)]{Fu02}
Y.~B. Fu and A.~Mielke.
\newblock A new identity for the surface-impedance matrix and its application
  to the determination of surface-wave speeds.
\newblock \emph{Proc. R. Soc. A}, 458\penalty0 (2026):\penalty0 2523--2543,
  2002.
\newblock \doi{10.2307/3067326}.

\bibitem[Kutsenko and Shuvalov(2013)]{Kutsenko12}
A.~A. Kutsenko and A.~L. Shuvalov.
\newblock Shear surface waves in phononic crystals.
\newblock \emph{J. Acoust. Soc. Am.}, 133\penalty0 (2):\penalty0 653--660,
  2013.
\newblock \doi{10.1121/1.4773266}.

\bibitem[Shuvalov(2003)]{Shuvalov03}
A.~L. Shuvalov.
\newblock A sextic formalism for three-dimensional elastodynamics of
  cylindrically anisotropic radially inhomogeneous materials.
\newblock \emph{Proc. R. Soc. A}, 459\penalty0 (2035):\penalty0 1611--1639,
  2003.

\bibitem[Hu et~al.(2012)Hu, Liu, and Bhattacharya]{Liu11}
L.~X. Hu, L.~P. Liu, and K.~Bhattacharya.
\newblock Existence of surface waves and band gaps in periodic heterogeneous
  half-spaces.
\newblock \emph{J. Elasticity}, 107\penalty0 (1):\penalty0 65--79, 2012.
\newblock \doi{10.1007/s10659-011-9339-0}.

\bibitem[Al'shits and Lothe(1981)]{Alshits81}
V.~I. Al'shits and J.~Lothe.
\newblock Comments on the relation between surface wave theory and the theory
  of reflection.
\newblock \emph{Wave Motion}, 3:\penalty0 297–--310, 1981.

\bibitem[Bai and Demmel(1998)]{Bai98}
Z.~Bai and J.~Demmel.
\newblock Using the matrix sign function to compute invariant subspaces.
\newblock \emph{SIAM J. Matrix Anal. Appl}, 19:\penalty0 205--225, 1998.

\end{thebibliography}
%\bibliographystyle{unsrtnat}%unsrt} plain}%uabbrvnat}%}%}%
%\end{document}

\end{document}